\documentclass[print, aps, prd, nofootinbib, ,superscriptaddress]{revtex4-2}
\usepackage{amssymb, amsmath, graphicx, epsfig, bm, appendix}
\usepackage{natbib}
\usepackage{graphicx}
\usepackage[caption=false]{subfig}
\usepackage{xcolor, hyperref}
\usepackage{slashed}
\usepackage{comment}
\usepackage{booktabs}

\definecolor{blue_refs}{rgb}{0., 0., 0.85}
\hypersetup{ 
colorlinks = true,
linkcolor = blue_refs,
citecolor = blue_refs,		
filecolor = blue_refs,		
urlcolor = blue_refs
}		
\renewcommand{\d}{\mathrm{d}}
\newcommand{\e}{\mathrm{e}}

\newcommand{\sT}{{\scriptscriptstyle T}}
\newcommand{\sL}{{\scriptscriptstyle L}}

\newcommand{\epss}{\varepsilon\!\!\!/}

\newcommand{\myparallel}{{\mkern3mu\vphantom{\perp}\vrule depth 0pt\mkern2mu\vrule depth 0pt\mkern3mu}}

\DeclareMathAlphabet{\pazocal}{OMS}{zplm}{m}{n}
\newcommand{\Q}{\pazocal{Q}}
\newcommand{\YQQ}{Y_{\Q\Q}}
\newcommand{\MQQ}{M_{\Q\Q}}
\usepackage[normalem]{ulem} % \sout{old text} for strikeout

\begin{document}
\title{Quark polarization and transverse momentum effects on double quarkonium production in hadronic collisions}

\author{Carlo Flore}
\email{carlo.flore@unica.it}
\affiliation{Dipartimento di Fisica, Università di Cagliari, Cittadella Universitaria, I-09042 Monserrato (CA), Italy}
\affiliation{INFN, Sezione di Cagliari, Cittadella Universitaria, I-09042 Monserrato (CA), Italy}

\author{Cristian Pisano}
\email{cristian.pisano@unica.it}
\affiliation{Dipartimento di Fisica, Università di Cagliari, Cittadella Universitaria, I-09042 Monserrato (CA), Italy}
\affiliation{INFN, Sezione di Cagliari, Cittadella Universitaria, I-09042 Monserrato (CA), Italy}

\begin{abstract}
We investigate the inclusive production of double quarkonia ($J/\psi$, $\psi(2S)$, $\Upsilon$ mesons) in polarized hadron-hadron collisions, considering the kinematic configuration where the transverse momentum of each pair of bound states is much smaller than its invariant mass.  Supported by nonrelativistic QCD arguments, we adopt the Color-Singlet Model for the quarkonium formation mechanism and assume the validity of transverse momentum dependent factorization. In strong analogy with dilepton production in the Drell-Yan processes, the azimuthal modulations of the cross section, calculated to the order $\alpha_s^4$ in the QCD coupling constant, can be expressed as convolutions of transverse momentum dependent distributions of light quarks and antiquarks inside the incoming hadrons. By adopting very recent parameterizations of these distributions, we show that a phenomenological study of these quantities in $\pi^-p\to J/\psi\,J/\psi\,X $ in the kinematic region covered by the COMPASS and AMBER fixed-target experiments at CERN, where the gluon contribution is found to be negligible, would provide direct access to the quark distributions. In particular, this will offer the possibility of a further sign change test of the quark Sivers function of the proton. The impact of our findings on similar studies about the gluon content of the proton in present and future fixed-target experiments at the LHC, like SMOG and LHCspin, is also demonstrated. 

\end{abstract}

\date{\today}
\maketitle

\section{Introduction}
Quarkonia (charmonia, bottomonia), {\it i.e.}~bound states of heavy quark-antiquark pairs $Q \overline Q$ ($c\bar c, b\bar b$), have played historically a fundamental role in the establishment of Quantum Chromodynamics (QCD) as the theory of strong interactions, mainly because of the clean signature they provide for different observables. From the theory point of view, the main simplification comes from the hierarchy  $M_Q\gg \Lambda_\text{QCD}$, with $M_Q$ being the heavy quark mass and $\Lambda_\text{QCD}$ the asymptotic scale parameter of QCD, such that $M_Q$ can be identified with the hard scale of the process, allowing for a perturbative expansion in the strong coupling constant $\alpha_s$. While the present theoretical frameworks all agree in providing a perturbative description of the creation of the $Q\overline Q$ pair, they differ in the treatment of the subsequent nonperturbative transition to the hadronic bound state.

In particular, the effective-field-theory approach of nonrelativistic QCD (NRQCD)~\cite{Bodwin:1994jh} establishes a separation of process-dependent short-distance coefficients, to be calculated perturbatively as an expansion in $\alpha_s$, from long-distance matrix elements (LDMEs). The former describe the production of a $Q \overline Q$ pair in a state $n =$ $^{2S+1}L_J^{[c]}$ with specific values of the orbital angular momentum $L$, spin $S$, total angular momentum $J$ and color configuration $c = 1, 8$, whereas the latter  encode the transition probability of the $Q\overline Q[n]$ configuration into the observed quarkonium state. The LDMEs are expected to be universal and have to be extracted from data or evaluated using nonperturbative techniques. Furthermore, scaling rules~\cite{Lepage:1992tx} predict that each of the LDMEs scales with a definite power of the relative velocity $v$ of the $Q\overline Q$ pair in the quarkonium rest frame, in the limit $v \ll 1$. Observables are therefore calculated, at fixed order, by means of a double expansion in $\alpha_s$ and $v$, with $v^2 \sim 0.3$ for charmonium and $v^2\sim 0.1$ for bottomonium. In the specific case of the production of a single $J/\psi$ meson,
according to the traditional Color-Singlet (CS) Model~\cite{Baier:1983va}, 
the $c\bar c$ pair is produced at short distances directly with the $J/\psi$ quantum numbers, namely as a $^3S^{[1]}_1$ state. To this channel NRQCD adds the leading relativistic corrections as well, given by the color-octet (CO) states $^1S_0^{[8]}$,  $^3S^{[8]}_1$ and $^3P^{[8]}_J$ with $J = 0, 1, 2$, up to the relative order ${\cal O}(v^4)$.  For $S$-wave quarkonia, the CS Model is then recovered in the limit $v \to 0$. Unfortunately, the present knowledge of the CO
LDMEs is not very accurate, because the different sets of their extracted values are not compatible with each other, even within the large uncertainties. Moreover, despite its success in explaining many experimental observations, NRQCD is not able to consistently account for all cross sections and polarization measurements of $J/\psi$ mesons produced both in proton-proton and in electron-proton collisions. For a recent review on this subject, see Ref.~\cite{Boer:2024ylx} and references therein. Although this complicates the use of quarkonia for precision studies, they remain very helpful tools to uncover new aspects of the structure of nucleons.

A striking example is provided by inclusive double $J/\psi$ production in proton-proton scattering at the LHC, which has been proposed as a way to probe the so-called transverse momentum dependent (TMD) parton distribution functions (or TMDs for short)~\cite{Lansberg:2017dzg,Scarpa:2019fol,Bor:2025ztq}. These  parton densities are receiving ever-growing attention from both the experimental and theoretical communities, mainly because they encode essential information on the transverse motion of partons inside nucleons, as well as their spin-orbit correlations. As such, they parameterize highly nontrivial features of the partonic structure of the proton. For instance, the Sivers function~\cite{Sivers:1989cc,Sivers:1990fh}, which describes the azimuthal distributions of unpolarized partons inside a proton transversely polarized with respect to its momentum, provides an indication on how much quarks and gluons contribute to the proton spin through their orbital angular momentum. At collider energies, double $J/\psi$ production is mainly sensitive to the so-far poorly known gluon TMDs. Indeed, the first extraction of the unpolarized gluon TMD has been performed using LHCb data at 13 TeV~\cite{LHCb:2016wuo}. The LHCb Collaboration has also measured the azimuthal modulations of one of the $J/\psi$ mesons in the Collins-Soper frame~\cite{LHCb:2023ybt}, which are generated by the distribution of linearly polarized gluons. All the available theoretical analyses have been performed at order $\alpha_s^4$ and, according to the NRQCD scaling rules, should not to be affected by the uncertainties due the CO LDMEs. Indeed, as shown in Ref.~\cite{He:2015qya}, the mixed CS-CO and CO-CO channels for double $J/\psi$ production are suppressed by factors of at least ${\cal O}(v^3)$ and ${\cal O}(v^6)$, respectively, as compared to the CS-CS contribution we are interested in. 

The suppression of the CO production mechanism is essential for TMD factorization to be applicable in hadron-hadron collisions. If the final $Q\overline Q$ pair is produced in a purely colorless state, only initial-state interactions (ISIs) between the active partons and the spectators can occur. Formally, these ISIs are encoded in past-pointing gauge links (Wilson lines) in the definition of the TMDs.  However, if colored $Q\overline Q$ pairs are produced as well, the combined effect of final state interactions (FSIs) and ISIs can lead to the breaking of factorization~\cite{Rogers:2010dm}. A second requirement for the applicability of the TMD approach is given by the presence of two well-separated scales: a nonperturbative one, sensitive to the intrinsic parton transverse momenta, and a hard one, which allows for a perturbative treatment. Hence one has to consider the kinematic region in which the transverse momentum of the $J/\psi$-pair (the nonperturbative scale) is much smaller than its invariant mass (the hard scale).

Motivated by the recent measurement of the transverse momentum distribution of $J/\psi$-pairs in pion-proton collisions released by the COMPASS Collaboration~\cite{COMPASS:2022djq}, in this paper we calculate for the first time, at the perturbative order $\alpha_s^4$ and within the combined frameworks of TMD factorization and CS production mechanism, the quark-antiquark annihilation channel of the reaction $h_1\, h_2\to {\cal Q}\, {\cal Q}\, X$, where $h_1, h_2$ are two (polarized) spin-1/2 hadrons and ${\cal Q}$ is a $S$-wave, spin-1 quarkonium state, {\it e.g.}\ a $J/\psi$, $\psi(2S)$, or $\Upsilon$ meson. More specifically, we provide the analytic expressions of the azimuthal modulations arising from the different convolutions of the quark and antiquark TMDs. Our study is therefore complementary to the ones in Refs.~\cite{Lansberg:2017dzg,Scarpa:2019fol,Bor:2025ztq}, in which only the gluon-gluon fusion channel has been taken into account. 

The present analysis should play an important role  also in the verification of the process dependence of TMDs, a property which is strongly related to their gauge link structure. One
of its most distinct consequences, still awaiting clear experimental support, is the expected sign change of the Sivers function in the Drell-Yan process with respect to semi-inclusive deep inelastic scattering (SIDIS)~\cite{Collins:2002kn}.  This is originated by the ISIs in the former reaction, leading to past pointing gauge links (as expected in double quarkonium production) as opposite  to the FSIs in SIDIS, generating future-pointing gauge links instead.  Obtaining a consistent picture among all these processes is therefore of fundamental importance for our understanding  of the TMD formalism and nonperturbative QCD in general.

The remainder of this paper is organized as follows. In Section~\ref{sec:qq-correlator} we  provide the definition of the leading-twist quark TMDs in terms of
QCD operators through a parametrization of the quark-quark correlator. Details of the calculation of the cross section for inclusive double quarkonium production in the TMD formalism are illustrated in Section~\ref{sec:di-onium}. Our results for the angular structure of the cross section are presented in Section~\ref{sec:azimuthal-modulations}. Phenomenological predictions for different fixed-target experiments are given in Section~\ref{sec:pheno}. Summary and conclusions are gathered in Section~\ref{sec:conclusions}. 

\section{\label{sec:qq-correlator} Quark-quark TMD correlator}

The transverse  momentum dependent quark-quark correlator for a spin-1/2 hadron is defined as  the Fourier transform of a nonlocal, forward matrix element of two quark fields between hadronic states  with momentum $P$, mass $M_h$ and vector spin $S$.  
More specifically, at leading twist, it is given by~\cite{Boer:1997nt}
\begin{align}
\Phi^{q\, [{\cal U}]}(x , \bm k_\sT)
& =  {\int}\frac{\d(\lambda{\cdot}P)\,\d^2\lambda_\sT}{(2\pi)^3}\ e^{ik\cdot\lambda}\,
\langle P, S |\,\overline\psi(0)\,{\cal U}_{[0,\lambda]}\,
\psi(\lambda)\,|P, S\rangle\,\big\rfloor_{\text{LF}}\,, 
\label{eq:corr-q}
\end{align}
where the nonlocality is restricted to the light-front (LF): $\lambda{\cdot}n\,{\equiv}\,0$, with  $n$ being  a light-like vector, $n^2 = 0$,
conjugate to $P$. Furthermore, the variables $x$ and $\bm k_\sT$ are defined through the Sudakov decomposition of the quark momentum emitted and reabsorbed by the hadron,
\begin{align}
k^\mu = x P^\mu \, + \, \frac{k^2 + \bm k_\sT^2}{2x\,  P\cdot n}\, n^\mu\,+\, k^\mu_\sT\,,     
\label{eq:k-sud}
\end{align}
with $\bm k_\sT^2 = -k_\sT^2$. In analogy to Eq.~\eqref{eq:k-sud}, the spin vector $S$ can be decomposed as
\begin{align}
S^\mu = \frac{S_\sL}{M_h}\left ( P^\mu - \frac{M_h^2}{P\cdot n}\,n^\mu \right ) + S_\sT^\mu\,,
\label{eq:Gamma-L}
\end{align}
with $S_\sL^2 + \bm S_\sT^2=1$. The process-dependent gauge link ${\cal U}_{[0,\lambda]}$ in Eq.~\eqref{eq:corr-q} sums up multiple gluon exchange contributions and makes the correlator gauge invariant. For the reaction under study, where only ISIs are present, the gauge link is a staple-like, past-pointing Wilson line. 

According to the hadron spin and omitting the dependence on the gauge link, the quark-quark correlator can be split into three parts: the unpolarized ($U$), the longitudinal polarized ($L$) and the transversely polarized ($T$) components,
\begin{align}
\Phi^q (x,\bm k_\sT ) =  \Phi^q_{U}(x,\bm k_\sT ) + \Phi^q_L(x,\bm k_\sT ) + \Phi^q_{T}(x,\bm k_\sT )  \,,
\label{eq:Gamma-parts}
\end{align}
which, at leading twist, can be parametrized in terms of quark TMD distributions as follows
\begin{align}
\Phi^q_U(x, \bm k_\sT) & = \frac{1}{2}\, \left  \{ f_1^q (x, \bm k_\sT^2) \slashed{P}\, + \, i  h_1^{\perp\, q}(x, \bm k_\sT^2)\frac{[\slashed{k}_\sT, \slashed{P}]}{2M_h}\right \} \,,  \nonumber \\
\Phi^q_L(x, \bm k_\sT) & = \frac{1}{2}\, S_L \left \{ g^q_{1\sL}(x, \bm k_\sT^2)\gamma^5 \slashed{P}\, + \, h^{\perp q}_{1\sL}(x, \bm k_\sT^2)\, \gamma^5 \, \frac{[\slashed{k}_\sT, \slashed{P}]}{2M_h} \right \} \,, \nonumber \\
\Phi^q_T(x, \bm k_\sT) & = \frac{1}{2}\,\bigg\{  - \frac{\epsilon^{k_\sT S_\sT}}{M_h}\, f^{\perp\,q}_{1\sT} (x, \bm k_\sT^2)\slashed{P} \,-\, 
    \frac{k_\sT \cdot S_\sT}{M_h}\,g_{1\sT}^{q}(x, \bm k_\sT^2)\, \gamma^5 \slashed{P}\,+ \, h_{1\sT}^{q}(x, \bm k_\sT^2)\,  \gamma^5 \,\frac{[\slashed{S}_\sT, \slashed{P}]}{2} \nonumber \\
    & \qquad \qquad - \frac{k_\sT \cdot S_\sT}{M_h} \, h^{\perp\, q}_{1\sT}(x, \bm k_\sT^2)\, \gamma^5 \, \frac{[\slashed{k}_\sT, \slashed{P}]}{2M_h} \bigg \} \,,
    \label{eq:Phi-par}
\end{align}
where we have introduced the antisymmetric transverse projector $\epsilon_\sT^{\mu\nu} = \epsilon_\sT^{\alpha\beta\mu\nu} P_\alpha n_\beta/P{\cdot}n$, with $\epsilon_\sT^{1 2} = +1$, as well as the notation
$\epsilon^{a b}_{\sT} \equiv \epsilon^{\alpha\beta}_{\sT} a_{\alpha} \,b_{\beta}$.
In Eq.~\eqref{eq:Phi-par}, the transverse momentum dependent unpolarized and helicity distributions are denoted by $f_1^q$ and $g^q_{1\sL}$, respectively. Moreover, we define the transversity distribution by the combination
\begin{align}
h_1^q (x, \bm k_\sT^2)\equiv h^q_{1\sT}(x, \bm k_\sT^2)\,+ \, \frac{\bm k_\sT^2}{2M^2_h}\,   h_{1\sT}^{\perp\,q} (x, \bm k_\sT^2) \,. 
\end{align} 
Of special importance are the time-reversal odd (T-odd) Sivers function $f_{1\sT}^{\perp\, q}$ and Boer-Mulders function $h_1^{\perp\,q}$~\cite{Boer:1997nt} as they can generate very interesting spin and azimuthal asymmetries in semi-inclusive processes. In particular, $h_1^{\perp\,q}$ gives rise to the $\cos 2\phi$ double Boer-Mulders modulation in the Drell-Yan process and to a violation of the Lam-Tung relation~\cite{Boer:1999mm,Boer:2002ju}.

Similarly, the correlator for antiquarks is defined as
\begin{align}
\overline \Phi^{q\,[{\cal U}]}(x , \bm k_\sT)
& =  -{\int}\frac{\d(\lambda{\cdot}P)\,\d^2\lambda_\sT}{(2\pi)^3}\ e^{-i k\cdot\lambda}\,
\langle P |\,\overline\psi(0)\,{\cal U}_{[0,\lambda]}\,
\psi(\lambda)\,|P\rangle\,\big\rfloor_{\text{LF}}\, ,
\label{eq:corr-qbar}
\end{align}
and can be decomposed in terms of antiquark TMD distributions in analogy to Eq.~\eqref{eq:Phi-par}. 

Finally, we note that the definitions in Eqs.~\eqref{eq:corr-q} and \eqref{eq:corr-qbar} refer to the so-called {\it unsubtracted} correlators. Once the ultraviolet and rapidity divergences are regularized, TMDs become dependent on two additional variables: the renormalization scale $\mu$ and the Collins-Soper scale $\zeta$. Such dependences are governed by QCD evolution equations~\cite{Collins:2011zzd,Boussarie:2023izj}. These scale are typically taken to be equal to the hard scale of the process under consideration.

\section{\label{sec:di-onium} Outline of the calculation} 

We study quarkonium-pair production in inelastic collisions of two spin-1/2 hadrons,
\begin{align}
h_1(P_1,S_1) + h_2(P_2,S_2) \to \Q(K_1) + \Q(K_2) + X  \,,
\end{align}
where the four-momenta of the particles are given within brackets, as well as the spin vectors $S_1$ and $S_2$ of the two incoming hadrons, which fulfill the relations $S_1^2 = S_2^2=-1$
and $S_1 \cdot P_1 = S_2 \cdot P_2 = 0$. The two quarkonia are taken to be almost back to back in the plane perpendicular to the directions of $h_1$ and $h_2$. Moreover, we assume that each of them is formed by a heavy quark-antiquark pair ($Q\overline Q$) directly produced in a Fock state with four-momentum $K_i$ (with $i=1,2$),  spin $S=1$, orbital angular momentum $L=0$, total angular momentum $J=1$,  and colorless configuration $c=1$. Namely, 
\begin{align}
\Q = Q\overline Q[^3S_1^{[1]}] \,.   
\end{align}
The squared invariant mass of each resonance is $M_{\Q}^2 = K_i^2$, where $M_\Q$ is twice the heavy quark mass up to small relativistic corrections. 

It has been shown in several publications~\cite{Echevarria:2019ynx,Fleming:2019pzj,Boer:2023zit} that, in order to describe quarkonium production at small transverse momentum, the TMD expression of the cross section for quarkonium production requires, instead of the usual LDMEs of NRQCD, the so-called {\it shape function}, combining soft gluon radiation and the formation of the bound state. As such, it encodes final state smearing effects, implying that the transverse momenta of the outgoing quarkonium state and the $Q \overline Q$ pair are different. Since in the CS production mechanism, each $Q\overline Q$ pair emerges from the hard scattering directly in a color-singlet configuration, with the same quantum numbers of the observed quarkonia ${\cal Q}$, soft gluon radiation and the corresponding smearing effects should be negligible and will not be considered in our analysis.

At the lowest order (LO) in the strong coupling constant $\alpha_s$, the underlying partonic processes are 
\begin{align}
q (k_1) + \bar q (k_2) \to \Q (K_1) + \Q (K_2) \quad\text{and}\quad g(k_1) + g (k_2) \to \Q (K_1) + \Q (K_2)\,,
\end{align}
where quarkonia are directly created from $q\bar q$ annihilation and $gg$ fusion, without the emission of additional partons. In the following, we focus only on the former contribution, since the latter has already been investigated widely within the TMD factorization approach~\cite{Lansberg:2017dzg,Scarpa:2019fol} using the scattering amplitude for the partonic process $gg\to J/\psi\,J/\psi$ calculated in Ref.~\cite{Qiao:2009kg}. Incidentally, we notice that, within the collinear factorization framework and in the CS Model, the gluon-gluon channel has also been studied at the order $\alpha_s^5$~\cite{Sun:2014gca}, whereas the quark-antiquark channel has been investigated in pion-nucleon scattering in fixed-target experiments at the order $\alpha_s^4$~\cite{Kartvelishvili:1983lrw,Humpert:1983qt,Ecclestone:1982yt}.

\subsection{\label{sec:kinematics} Kinematics}

In order to perform a Sudakov decomposition of the particle momenta, we introduce two light-like vectors $n$ and $\overline n$, with  $n^2 = \overline n^2 =0$ and $n\cdot \overline n =1$, such that the light-cone components of every vector $v$ are defined as $v^+ \equiv v \cdot \overline n$ and $v^-\equiv v\cdot n$, while perpendicular vectors $v_\sT$ always refer to the components of $v$ orthogonal to both the momenta of the incoming hadrons, ${P}_1$ and ${P}_2$, with $v_\sT^2 = - \bm v_\sT^2$. 
Hence, for the initial hadron momenta we can write
\begin{align}
{P}_1^\mu & = {P}_1^+n^\mu + \frac{M_{h_1}^2}{2{P}_1^+}\, \overline n^\mu\,, \qquad
{P}_2^\mu  = \frac{M_{h_2}^2}{2{P}_2^-}\, n^\mu  + {P}_2^- \overline n^\mu\,,
\label{eq:P1-P2}
\end{align}
with $M_{h_1}$, $M_{h_2}$ being the hadron masses and $s\approx 2 P_1^+ P_2^-$. The parton momenta can be expressed in terms of the light-cone momentum fractions $(x_1, x_2)$ and the intrinsic transverse momenta $(k_{1\sT}, k_{2\sT})$ as
\begin{align}
k_{1}^\mu & = x_1 P_1^+ n^\mu + \frac{k_1^2 + \bm k_{1\sT}^2}{2 x_1 P_1^+}\, \overline n^\mu +  k_{1\sT}^\mu\,, \qquad
k_{2}^\mu  =  \frac{k_2^2 + \bm k_{2\sT}^2}{2 x_2 P_2^-}\, n^\mu + x_2 {P}_2^- \overline n^\mu + k_{2\sT}^\mu\,,
\end{align}
whereas for the final quarkonium momenta we have 
\begin{equation}
    K_1^\mu =   K_1^+  {n}^\mu +  \frac{M_{\cal Q}^2 + \bm K_{1\perp}^2}{2K_1^+}\, \overline{n}^\mu + K_{1\perp}^\mu\,, \qquad K_2^\mu =  K_2^+\, {n}^\mu + \frac{M_{\cal Q}^2 + \bm K_{2\perp}^2}{2 K_2^+} \overline{n}^\mu + K_{2\perp}^\mu\, .
       \label{eq:K1-K2}
\end{equation}
The momentum-conserving delta function can therefore be decomposed as follows 
\begin{align}
\delta(k_1 + k_2 - K_1-K_2) & \approx \frac{2}{s}\delta \left (x_1  - \frac{K_1^+ + K_2^+}{P_1^+} \right ) \, \delta \left (x_2 - \frac{1}{2 P_2^-} \left (\frac{M_{1\perp}^2}{K_1^+} + \frac{M_{2\perp}^2}{K_2^+} \right )  \right )\,  \delta^2(\bm k_{1\sT} + \bm k_{2\sT} - \bm q_\sT) \, ,
\label{eq:delta}
\end{align}
with $M_{i\perp}^2 = M_\Q^2 + \bm K_{i\perp}^2$, which fixes $x_1$ and $x_2$ as 
\begin{equation}
    x_1 = \frac{K_1^+ + K_2^+}{P_1^+} \,,  \qquad 
    x_2 =  \frac{1}{2 P_2^-} \left (\frac{M_{1\perp}^2}{K_1^+} + \frac{M_{2\perp}^2}{K_2^+} \right ) \, .
\label{eq:x1-x2}
\end{equation}
In particular, in the hadronic center-of-mass (c.m.) frame, $K_1^+$ and $K_2^+$ can be expressed in terms of the rapidities $y_1$ and $y_2$ of the two final quarkonium states:
\begin{align}
K_1^+ = \frac{M_{1 \perp}}{\sqrt{2}}\, \,e^{y_1}\,, \qquad K_2^+ =  \frac{M_{2 \perp}}{\sqrt{2}}\, \,e^{y_2}\,.  
\end{align}

Our final results can be expressed in terms of the variables
\begin{align}
    z_1 = \frac{K_1 \cdot k_1}{k_1\cdot k_2} = \frac{1}{1 + e^{y_1 - y_2}}\,,\qquad 
    z_2 = \frac{K_2 \cdot k_1}{k_1\cdot k_2} = \frac{1}{1 + e^{y_2 - y_1}}\,.
    \label{eq:z1-z2}
\end{align}
Because of momentum conservation, $k_1 + k_2 = K_1 + K_2$, one has $z_1 + z_2 = 1$;  we can thus define
\begin{equation}
z \equiv z_1, \qquad 1-z \equiv z_2\,.
\label{eq:z}
\end{equation}
Finally, as it will be useful later, we introduce the variable
\begin{equation}
    \YQQ = \frac12 \ln\left({\frac{x_1}{x_2}}\right) \simeq y_1 +  \frac12 \ln\left({\frac{z}{1-z}}\right) = \frac{y_1+y_2}{2}\,,
\label{eq:Ypsipsi}
\end{equation}
that is the quarkonium-pair rapidity in the c.m.~frame.
%%%%%%%%%%%%%%%%%%%%%%%%%%%%%%%%%%%%%%%%%%%%%%%%
\begin{figure*}[t]
\centering
\includegraphics[width=9.5cm, keepaspectratio]{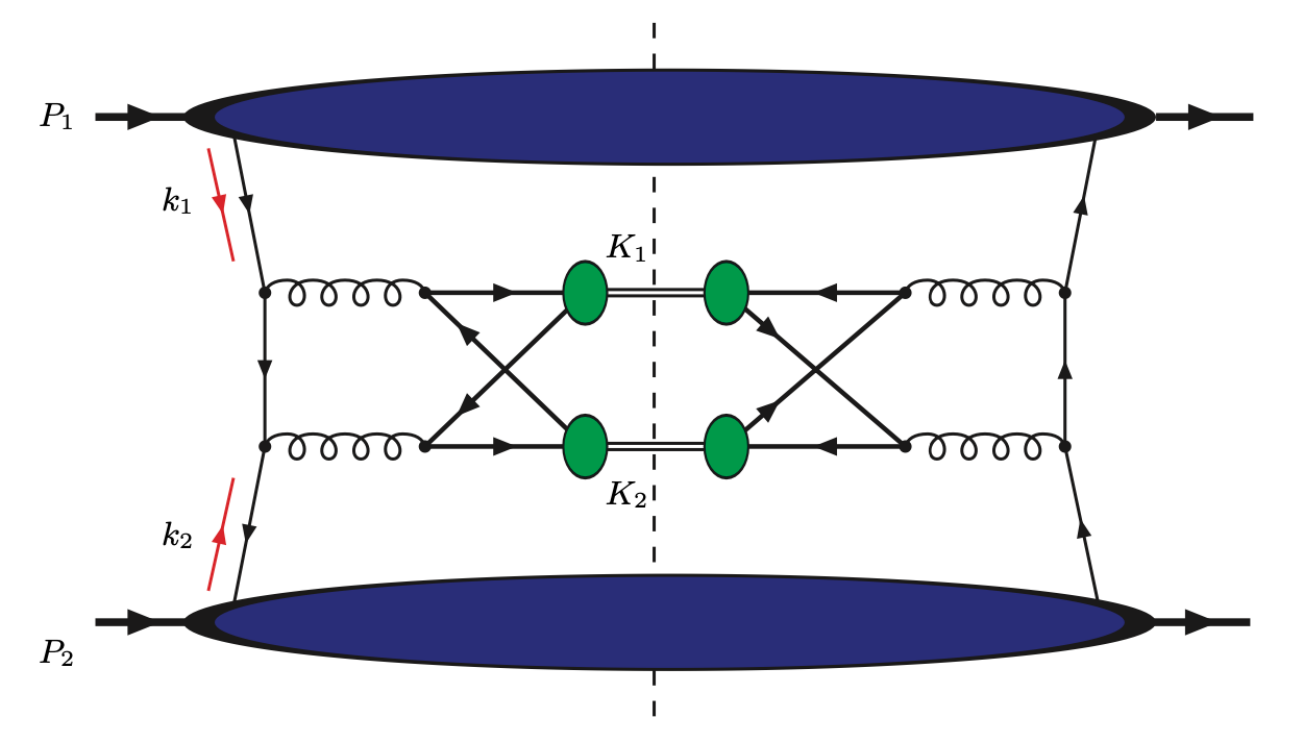}
\caption{Representative cut diagram for the process $h_1(P_1) + h_2(P_2) \to \Q(K_1) + \Q(K_2) + X$ through  quark-antiquark annihilation at LO. The upper and lower blue ovals describe the correlators $\Phi^q(x_1, \bm k_{1\sT}^2)$ and $\overline\Phi^q(x_2, \bm k_{2\sT}^2)$, respectively.} 
\label{eq:sq-amplitude}
\end{figure*}
%%%%%%%%%%%%%%%%%%%%%%%%%%%%%%%%%%%%%%%%%%%%%%%%%

\subsection{\label{sec:cross-section} General expression of the TMD cross section}

Assuming TMD factorization, the cross section in its most differential form, schematically described by the Feynman cut diagram in Fig.~\ref{eq:sq-amplitude}, reads
\begin{align}
\d\sigma & = \frac{1}{2s}\, \frac{\d^3 K_1}{(2\pi)^32 E_1}\, \frac{\d^3 K_2}{(2\pi)^3 2E_2}\int \d x_1 \, \d x_2  \, \d^2 k_{1\sT}\, \d^2 k_{2 \sT}  \, (2 \pi)^4 \, \delta^4(k_1 + k_2 - K_1-K_2) \nonumber\\
& \qquad\qquad \qquad  
\times \sum_q \Phi^q(x_1, \bm k_{1\sT})  
\otimes \overline \Phi^q(x_2, \bm k_{2\sT})\,\otimes \vert {\cal M}_{q\, \bar q\to \Q\, \Q} (k_1, k_2, K_1, K_2)\vert^2 \, + \{\Phi^q \leftrightarrow \overline{\Phi}^q \} \,,
\label{eq:PS}
\end{align}
where the sum runs over the (anti)quark flavors, the symbol $\otimes$ indicates that a trace over the Dirac indices is taken and ${\cal M}_{q\, \bar q\to \Q\, \Q}$ is the scattering amplitude for the process ${q\, \bar q\to {\cal Q}\, {\cal Q}}$. 

At this point it is convenient to introduce the sum and difference of the quarkonium transverse momenta, $K_\perp = (K_{1\perp} - K_{2\perp})/2$ and $q_\sT = K_{1\perp} + K_{2\perp}$ with $\vert q_\sT\vert \ll \vert K_\perp\vert$ because of the back-to-back configuration we are considering. We can therefore use the approximate transverse momenta $K_{1\perp} \approx K_{\perp}$ and $K_{2\perp} \approx -K_{\perp}$, implying $M_{1\perp}^2 \approx M_{2\perp}^2 \approx M_\perp^2 = M_\Q^2 + \bm K_\perp^2$. The cross section in Eq.~\eqref{eq:PS} can therefore be written in the following form:
\begin{align}
\frac{\d\sigma}{\d y_1\, \d y_2\,  \d^2 \bm{K}_\perp\, \d^2 \bm{q}_\sT}  & = \frac{1}{16 \pi^2 s^2 } \, \int \d^2 k_{1\sT}\, \d^2 k_{2 \sT}\,\delta^2(\bm k_{1\sT} + \bm k_{2\sT} - \bm q_\sT) \label{eq:cs} \nonumber \\ 
& \qquad \quad \times \sum_{q} \,\Phi^q(x_1, \bm k_{1\sT})  \otimes \overline \Phi^q(x_2, \bm k_{2\sT})\,\otimes \vert {\cal M}_{q\, \bar q \to \Q\, \Q}\vert^2 \, + \{\Phi^q \leftrightarrow \overline{\Phi}^q \}\,,
\end{align}
where the momentum fractions $x_1$ and $x_2$, given in Eq.~\eqref{eq:x1-x2}, can be approximated as
\begin{equation}
    x_1 \simeq \frac{M_\perp}{\sqrt{s}\,(1-z)}\,e^{y_1},\qquad x_2 \simeq \frac{M_\perp}{\sqrt{s}\,z}\,e^{-y_1}\, .
    \label{eq:x1-x2-MT-z}
\end{equation}
Furthermore, we note that using Eq.~\eqref{eq:x1-x2-MT-z} and the definition of $\hat{s} \equiv (k_1+k_2)^2= x_1 x_2 s = \MQQ^2\,$, one gets
\begin{equation}
\MQQ^2 = \frac{M_\perp^2}{z\,(1-z)}\,,
\label{eq:Mpsipsi-MT-z}
\end{equation}
which relates the invariant mass of the quarkonium pair with the transverse mass of one of the produced quarkonia. 

In order to obtain an expression for the cross section in terms of parton distributions, we have to insert the parametrization of the TMD quark correlator $\Phi^q$, given in Eq.~\eqref{eq:Phi-par}, into Eq.~\eqref{eq:cs}. The parametrization of $\overline \Phi^q$ can be obtained from Eq.~\eqref{eq:Phi-par} by performing the replacements $f^q \to f^{\bar q}$ for all TMDs, except for $g_{1\sL}^q$ and  $g_{1\sT}^{\perp \,q}$ since their antiquark counterparts acquire a minus sign due to the specific properties under charge conjugation of the corresponding Dirac operators~\cite{Tangerman:1994eh}, that is $g_{1\sL}^q\to - g_{1\sL}^{\bar q}$ and $g_{1\sT}^{q}\to - g_{1\sT}^{\bar q}$. Moreover,  we point out that for the hadron with momentum $P_2$ the roles of the forward and backward light-cone directions are exchanged as compared to the other hadron with momentum $P_1$, hence in the parametrization of $\overline \Phi^q (x_2, \bm k_{2\sT})$ the epsilon tensor should be taken with opposite sign with respect to 
$\Phi^q (x_1, \bm k_{1\sT})$: $\epsilon_\sT^{\mu\nu} \to -\epsilon_\sT^{\mu\nu}$. The details of the calculation of the scattering amplitude ${\cal M}_{q\, \bar q \to \Q\, \Q}$ can be found in Section~\ref{sec:amplitude}.

%%%%%%%%%%%%%%%%%%%%%%%%%%%%%%%%%%%%%%%%%%%%%%%%
\begin{figure}[t]
\centering
\subfloat[\label{fig:amplitude-a}]{
    \includegraphics[width = 6.75cm, keepaspectratio]{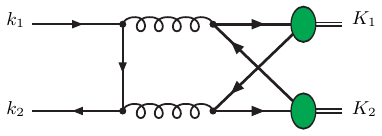}
    }
\hspace{.5cm}
\subfloat[\label{fig:amplitude-b}]{
    \includegraphics[width = 6.75cm, keepaspectratio]{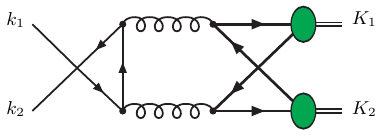}
    }
\caption{Feynman diagrams contributing to the scattering amplitude for the partonic process $q \,\overline q \to \Q \, \Q$ at order $\alpha_s^2$. The other two crossed diagrams, in which the directions of the arrows in the heavy (anti)quark lines are reversed, are not shown explicitly.}  
\label{fig:amplitude}
\end{figure}
%%%%%%%%%%%%%%%%%%%%%%%%%%%%%%%%%%%%%%%%%%%%%%%%%

\subsection{\label{sec:amplitude}{Scattering amplitude}}

The scattering amplitude for the process $q (k_1)\,+\, \bar q(k_2)\to \Q (K_1) + \Q (K_2)$, with ${\cal Q}$ being a $^3S_1$ quarkonium state, can be written in the form
\begin{align}
{\cal M} (k_1,k_2; K_1, K_2) = \frac{1}{16 \pi M_\Q}\, R_\Q^2(0)\, {\rm Tr}\left [ O^{\mu\nu} (k_1, k_2; K_1, K_2) (\slashed{K}_1 - M_\Q)\,\epss_{\lambda_{1} }^*(K_1)\,  \gamma_\nu (\slashed{K}_2 - M_\Q)\, \epss^*_{\lambda_2} (K_2)\gamma_\mu\right ] \,,
\label{eq:3S1a}
\end{align}
where $R_{\cal Q}(0)$ is the radial wave function at the origin, the trace is taken over the Dirac indices and $\epsilon_{\lambda_{1,2}} (K_i)$ are the polarization vectors of the two spin-1 quarkonia. In the non-relativistic limit, we neglect the relative momentum of the heavy-quark pair forming the bound state, the operator $O^{\mu\nu}(k_1, k_2; K_1, K_2)$ is thus calculated from the partonic subprocess
\begin{align}
q_m (k_1)\, + \, \bar q_n(k_2)\to 
Q_i (K_1/2) + \overline Q_j(K_2/2) + Q_k (K_1/2) + \overline Q_\ell(K_2/2) \,, 
\end{align}
{where the subscripts denote the color indices of the (anti)quark fields. At the order $\alpha_s^2$, $O^{\mu\nu}$ consists of four terms,
\begin{align}
O^{\mu\nu}(k_1, k_2; K_1, K_2) = \sum_{l =1}^4 \, O_{(l)}^{\mu\nu}  (k_1, k_2; K_1, K_2)\,.    
\end{align}
In particular, $O_{(1)}^{\mu\nu}$ can be obtained from the Feynman diagram in Fig.~\ref{fig:amplitude-a} by projecting the color with the operator $\delta_{i\ell}\delta_{jk}/N_c$, with $N_c$ being the number of colors, without including the Dirac spinors corresponding to the incoming light (anti)quarks
and the heavy quark legs, since they are absorbed into the
definitions of the (anti)quark correlators and of the spin projection operators inside the trace in Eq.~\eqref{eq:3S1a}, respectively. We find
\begin{align}
 O_{(1)}^{\mu\nu} (k_1, k_2; K_1, K_2) & = -32\, g_s^4\, t_{mn}^a t_{nr}^b t^b_{i j}\, t^a_{k\ell}\, \frac{1}{N_c}{\delta_{i\ell}}\, \delta_{j k}\,\frac{1}{(K_1+K_2)^4}\,\gamma^\mu \,\frac{2 \slashed{k}_1 - \slashed{K}_1 -\slashed{K}_2}{(2k_1 -K_1-K_2)^2}\, \gamma^\nu  \nonumber \\
 &  = -64\, g_s^4\, t_{mn}^a t_{nr}^b \frac{1}{2}\delta^{ab}\, \frac{1}{N_c}\, \frac{1}{(K_1+K_2)^4}\,\frac{\gamma^\mu (k_1-k_2)^\nu}{(k_1-k_2)^2}\,.
 \end{align}

 Similarly, from the diagram in Fig.~\ref{fig:amplitude-b}, we get
 \begin{align}
 O_{(2)}^{\mu\nu} (k_1, k_2; K_1, K_2) & = -32\, g_s^4\, t^b_{mn} t^a_{nr} \, t^b_{i j}\, t^a_{k\ell}\,  \frac{1}{N_c}{\delta_{i\ell}}\,\delta_{j k}\,\frac{1}{(K_1+K_2)^4}\,\gamma^\nu \,\frac{2 \slashed{k}_1 - \slashed{K}_1 -\slashed{K}_2}{(2k_1 -K_1-K_2)^2}\, \gamma^\mu \nonumber \\
 & =-64\, g_s^4\, t^b_{mn} t^a_{nr} \,\frac{1}{2}\delta^{ab}\,  \frac{1}{N_c}\, \frac{1}{(K_1+K_2)^4} \, \frac{\gamma^\nu ( k_1 - k_2)^\mu }{(k_1 - k_2)^2}\,.
\end{align}
From the crossed diagrams obtained from Fig.~\ref{fig:amplitude} by reversing the heavy (anti)quark lines, one finds $O^{\mu\nu}_{(3)} = O^{\mu\nu}_{(1)}$  and  $O^{\mu\nu}_{(4)} = O^{\mu\nu}_{(2)}$. After summing up all the contributions, the amplitude in Eq.~\eqref{eq:3S1a} can be written as~\cite{Kartvelishvili:1983lrw}
\begin{align}
 {\cal{M}} (k_1,k_2; K_1, K_2) & =  - \frac{1}{\pi\, M_\Q }\, 4\,  g_s^4\, t^a_{mn} t^a_{nr} \,  \frac{1}{2N_c}\,\frac{1}{(K_1+K_2)^4 (k_1-k_2)^2}\,R^2_{\cal Q}(0)\, D_{\mu\nu}R^{\mu\nu} \nonumber \\
 &  = - \frac{1}{\pi\, M_\Q}\, 2\,  g_s^4\,\frac{N_c^2-1}{4N_c}\, \delta_{mr} \,  \frac{1}{N_c}\,\frac{1}{(K_1+K_2)^4 (k_1-k_2)^2}\,R^2_{\cal Q}(0)\, D_{\mu\nu}R^{\mu\nu}\,, 
\end{align}
with 
\begin{align}
D_{\mu\nu} & =  {\rm Tr}\left [ (\slashed{K}_1 - M_{\Q})\,\epss_{\lambda_{1} }^*(K_1)\,  \gamma_\nu (\slashed{K}_2 - M_{\Q})\, \epss^*_{\lambda_2} (K_2)\gamma_\mu\right ] \,,\nonumber \\
R^{\mu\nu} & = \gamma^\mu ( k_1 - k_2)^\nu +  \gamma^\nu ( k_1 - k_2)^\mu \,.
\end{align}

We point out that our result for the unpolarized partonic cross section of the process $q \bar q \to {\cal Q}\,{\cal Q}$, 
\begin{align}
\frac{\d\sigma}{\d \hat t} = \frac{1}{16\pi\hat s^2}\, \overline{ \vert\widetilde{\cal M}}\vert ^2
\end{align}
with $\widetilde{\cal M} \equiv \overline{v} (k_2)\, {\cal M }\, u(k_1)$ and 
\begin{align}
\overline {\vert \widetilde{{\cal M }}\vert^2} = \frac{1}{N_c^2}\,\frac{1}{4}\, \sum_\text{colors} \, \sum_\text{spins} \vert { \widetilde{\cal M }}\vert^2 \,,  
\end{align}
expressed in terms of the Mandelstam variables $\hat s=(k_1+k_2)^2$ and $\hat t = (k_1-K_1)^2$, is in agreement with the one in Ref.~\cite{Kartvelishvili:1983lrw}. 

\section{\label{sec:azimuthal-modulations} Angular structure of the cross section}

We now present the analytical expression of the fully differential cross section, as obtained from Eq.~\eqref{eq:cs} after performing the calculations outlined in the previous section.
In a frame where the longitudinal direction is set by the momenta of the two incoming hadrons $h_1$ and $h_2$, we denote by $\phi_\sT$ and $\phi_\perp$ the azimuthal angles of $q_\sT$ and $K_\perp$, respectively. Analogously, $\phi_{S_1}$ and $\phi_{S_2}$ are the azimuthal angles of the spin vectors of $h_1$ and $h_2$. Thus the cross section
\begin{align}
\d\sigma \equiv  \frac{\d\sigma}{\d y_1\, \d y_2\,  \d^2 \bm{K}_\perp\, \d^2 \bm{q}_\sT}    
\label{eq;dsigma-def}
\end{align}
can be written as
\begin{align}
  \d \sigma  & =  \frac{131072}{243} \, \frac{\alpha_s^4}{M_\Q^2 \, M_\perp^6 \, s} \,\vert R_\Q(0)\vert^4\,  z^4 (1-z)^4 \, \bigg \{ F_{UU} + F_{UU}^{\,\cos2\left(\phi_\sT-\phi_\perp\right)} \cos2\left (\phi_\sT - \phi_\perp \right) \nonumber \\
   & \qquad + \, \bigg [ S_{1\sL} \, F_{LU}^{\,\sin 2 \left (\phi_\sT -\phi_\perp \right)}  \, + \,  S_{2\sL}\, F_{UL}^{\,\sin 2 \left (\phi_\sT -\phi_\perp \right )} \bigg ]\, \sin 2 (\phi_\sT -\phi_\perp) \,+\, \vert \bm S_{1\sT}\vert \bigg [ F_{TU}^{\sin(\phi_\sT-\phi_{S_1}) }\sin(\phi_\sT-\phi_{S_1}) \nonumber \\
  & \qquad +\,   F_{TU}^{\,\sin\left (\phi_\sT + \phi_{S_1} -2\phi_\perp \right )  }\sin \left (\phi_\sT +\phi_{S_1} - 2 \phi_\perp \right ) \,+ \,  F_{TU}^{\,\sin\left (3\phi_\sT - \phi_{S_1} -2\phi_\perp\right )}\sin \left (3\phi_\sT -\phi_{S_1} - 2 \phi_\perp \right ) \bigg ] \nonumber 
  \end{align}\\
  \begin{align}
  & \qquad +\, \vert \bm S_{2\sT}\vert \bigg [ F_{UT}^{\,\sin \left (\phi_\sT-\phi_{S_2} \right ) }\sin \left (\phi_\sT-\phi_{S_2}\right )  \,  +\,  F_{UT}^{\,\sin(\phi_\sT + \phi_{S_2} -2\phi_\perp )  }\sin \left (\phi_\sT +\phi_{S_2} - 2 \phi_\perp \right ) \nonumber \\
  & \qquad + \,  F_{UT}^{\,\sin\left (3\phi_\sT - \phi_{S_2} -2\phi_\perp \right)  }\sin\left (3\phi_\sT -\phi_{S_2} - 2 \phi_\perp \right)\bigg ]  \, +\,  S_{1\sL} \, S_{2\sL} \,\bigg [ F_{LL} \,+\,  F^{\, \cos2(\phi_\sT-\phi_\perp)}_{LL} \, \cos 2(\phi_\sT-\phi_\perp) \bigg ] \nonumber \\
  & \qquad +\, S_{1\sL} \vert \bm S_{2\sT} \vert\,\bigg [ F_{LT}^{\,\cos (\phi_\sT - \phi_{S_2})}\cos (\phi_\sT - \phi_{S_2}) \, + \,F_{LT}^{\,\cos (\phi_\sT + \phi_{S_2} - 2\phi_\perp)}\cos (\phi_\sT + \phi_{S_2} - 2 \phi_\perp) 
  \nonumber \\
  & \qquad + \, F_{LT}^{\,\cos (3\phi_\sT -\phi_{S_2} - 2\phi_\perp)}\cos (3\phi_\sT -\phi_{S_2} - 2 \phi_\perp)\bigg ]
    \, +\, \vert \bm S_{1\sT} \vert \,S_{2\sL} \,\bigg [ F_{TL}^{\,\cos (\phi_\sT - \phi_{S_1})}\cos (\phi_\sT - \phi_{S_1}) \, \nonumber \\
   & \qquad + \,F_{TL}^{\,\cos (\phi_\sT + \phi_{S_1} - 2\phi_\perp)}\cos (\phi_\sT + \phi_{S_1} - 2 \phi_\perp) \, + \, F_{TL}^{\,\cos (3\phi_\sT -\phi_{S_1} - 2\phi_\perp)}\cos (3\phi_\sT -\phi_{S_1} - 2 \phi_\perp)\bigg ] \nonumber \\
   & \qquad +\, \vert \bm S_{1\sT}\vert \vert \bm S_{2\sT}\vert \bigg [ F_{TT}^{\, \cos(\phi_{S_1} -\phi_{S_2})}\, \cos(\phi_{S_1} -\phi_{S_2}) 
   \nonumber \,+ \, F_{TT}^{\, \cos(2\phi_\sT -\phi_{S_1} -\phi_{S_2})}  \, \cos(2\phi_\sT -\phi_{S_1} -\phi_{S_2}) \nonumber \\
   & \qquad +\, F_{TT}^{\,\cos(\phi_{S_1} + \phi_{S_2}-2\phi_\perp)}\, \cos(\phi_{S_1} + \phi_{S_2}-2\phi_\perp)\,+\, F_{TT}^{\, cos(2\phi_\sT + \phi_{S_1} - \phi_{S_2}-2\phi_\perp)}\, \cos(2\phi_\sT + \phi_{S_1} - \phi_{S_2}-2\phi_\perp)\nonumber \\
   & \qquad +\,  F_{TT}^{\, cos(2\phi_\sT - \phi_{S_1} + \phi_{S_2}-2\phi_\perp)}\, \cos(2\phi_\sT - \phi_{S_1} +\phi_{S_2}-2\phi_\perp) \nonumber \\
   & \qquad + \, F_{TT}^{\, \cos(4\phi_\sT -\phi_{S_1} - \phi_{S_2} -2\phi_\perp)}\,\cos(4\phi_\sT -\phi_{S_1} - \phi_{S_2} -2\phi_\perp) \bigg ]\,,
  \label{eq:dsigma}
\end{align}
where the subscripts of the structure functions $F$ refer to the polarizations of the incoming hadrons, while the superscripts specify the corresponding azimuthal modulation. Each structure function can be factorized into a (perturbatively calculable) hard part, and a convolution of TMDs, which in  momentum space is defined as
\begin{align}
    {\cal C} [w\, f^q_1\, f^{\bar q}_2] & 
    \equiv {\cal C} [w(\bm k_{1\sT}, \bm k_{2\sT})\, f^{q/h_1}_1(x_1,\bm k_{1\sT})\, f^{\bar q/h_2}_2(x_2,\bm k_{2\sT})] \, +\, \{ q \leftrightarrow \bar q \}\nonumber \\ 
    &  = \int \d^2\bm k_{1\sT} \,\d^2\bm k_{2\sT}\, w(\bm k_{1\sT}, \bm k_{2\sT})\, f^{q/h_1}_1(x_1,\bm k_{1\sT})\, f^{\bar q/h_2}_2(x_2,\bm k_{2\sT})\, \delta^{2}(\bm k_{1\sT} + \bm k_{2\sT} - \bm q_\sT) \, + \{ q \leftrightarrow \bar q \}\, ,
\end{align}
where $f_i$, with $i=1,2$, are the TMDs of hadrons $h_i$ and $w(\bm p_{1\sT}, \bm p_{2\sT})$ is a proper weight function that depends on the particular  distributions involved. A sum over the quark flavors is understood. Explicitly, if we introduce the hard functions 
\begin{align}
& H \left ( z, \frac{M^2_\Q}{M^2_\perp} \right )  =    5 -  12\, z (1-z)\,\left (  1-\frac{M_{\cal Q}^2}{M_\perp^2}\right ) - \frac{M_{\cal Q}^2}{M_\perp^2} \,,\nonumber\\[2mm] 
& \Delta H \left ( z, \frac{M^2_\Q}{M^2_\perp} \right )  = -\left (  1-\frac{M_{\cal Q}^2}{M_\perp^2}\right )\left [ 1- 12\, z (1-z) \right ]\,, 
\label{eq:hard-f}
\end{align}
we find
\begin{align}
F_{UU}  & =H \left ( z, \frac{M^2_\Q}{M^2_\perp} \right ) \, {\cal C}[f_1^{q} f_1^{\bar q}] 
\,, \label{eq:F_UU}
\\
F_{UU}^{\,\cos 2 (\phi_\sT-\phi_\perp)} & = \Delta H \left ( z, \frac{M^2_\Q}{M^2_\perp} \right )\,{\cal C}[w_{UU}\,h_1^{\perp\, q}\, h_1^{\perp\, \bar q}]  
\,,\label{eq:F_UU_cos2phi}
\end{align}
\begin{align}%\\
F_{LU}^{\,\sin 2 (\phi_\sT-\phi_\perp)} & =- \Delta H \left ( z, \frac{M^2_\Q}{M^2_\perp} \right ) \,{\cal C}[w_{LU}\, h_{1\sL}^{q}\, h_{1}^{\perp\, \bar q}] 
\,,\label{eq:F_UU_sin2phi}
\\
F_{TU}^{\,\sin\left (\phi_\sT-\phi_{S_1} \right ) } & = - H \left ( z, \frac{M^2_\Q}{M^2_\perp} \right )\,
{\cal C} [w^1_{TU}\, f_{1\sT}^{\perp\, q}\, f_1^{\bar q} ]
\,,\\
F_{TU}^{\,\sin \left (\phi_\sT + \phi_{S_1} - 2 \phi_\perp \right )} & = - \Delta H \left ( z, \frac{M^2_\Q}{M^2_\perp} \right )\,{\cal C}[w_{TU}^{2}\,h_1^{q}\, h_1^{\perp\, \bar q}]  \,, \\
F_{TU}^{\, \sin \left (3 \phi_\sT - \phi_{S_1} - 2 \phi_\perp \right )} & =  - \Delta H \left ( z, \frac{M^2_\Q}{M^2_\perp} \right )\,{\cal C}[w^3_{TU}\, h_{1 \sT}^{\perp\, q}\, h_1^{\perp\, \bar q}]  \,,\\
F_{UL}^{\,\sin 2 (\phi_\sT-\phi_\perp)} & =   \Delta H \left ( z, \frac{M^2_\Q}{M^2_\perp} \right ) \,{\cal C}[w_{UL}\,h_{1}^{\perp\, q}\, h_{1\sL}^{\bar q}] 
\,, \\
F_{UT}^{\,\sin\left (\phi_\sT-\phi_{S_2} \right ) } & = H \left ( z, \frac{M^2_\Q}{M^2_\perp} \right )\,
{\cal C} [w^1_{UT}\,  f_1^q\, f_{1\sT}^{\perp\, \bar q}\,]
\,, \label{eq:F_UT-Sivers}\\
F_{UT}^{\,\sin \left (\phi_\sT + \phi_{S_2} - 2 \phi_\perp \right )} & = \Delta H \left ( z, \frac{M^2_\Q}{M^2_\perp} \right ) \,{\cal C}[w_{UT}^{2}\,h_1^{\perp\, q}\, h_1^{\bar q}]  \,, \\
F_{UT}^{\, \sin \left (3 \phi_\sT - \phi_{S_2} - 2 \phi_\perp \right )} & = \Delta H \left ( z, \frac{M^2_\Q}{M^2_\perp} \right ) \,{\cal C}[w^3_{UT}\,h_1^{\perp\, q}\, h_{1 \sT}^{\perp\, \bar q}]  \,,\\
F_{LL}  & = - H \left ( z, \frac{M^2_\Q}{M^2_\perp} \right )\, {\cal C}[g_{1\sL}^{q} \, g_{1\sL}^{\bar q}]  \,, \\
F_{LL}^{\,\cos 2 (\phi_\sT-\phi_\perp)} & = \Delta H \left ( z, \frac{M^2_\Q}{M^2_\perp} \right ) \,{\cal C}[w_{LL}\,h_{1\sL}^{\perp\, q}\, h_{1 \sL}^{\perp\, \bar q}]  
\,,\\
F_{LT}^{\,\cos (\phi_\sT - \phi_{S_2})} & = -  H \left ( z, \frac{M^2_\Q}{M^2_\perp} \right )\, {\cal C}[ w^1_{LT}\, g_{1\sL}^{q} \, g_{1\sT}^{\bar q}] \,, \\
F_{LT}^{\,\cos (\phi_\sT + \phi_{S_2} - 2 \phi_\perp)} & = \Delta H \left ( z, \frac{M^2_\Q}{M^2_\perp} \right )\, {\cal C}[ w^2_{LT}\, h_{1\sL}^{\perp\, q} \, h_{1}^{\bar q}]\,, \\
F_{LT}^{\,\cos (3\phi_\sT -\phi_{S_2} - 2\phi_\perp)} & = \Delta H \left ( z, \frac{M^2_\Q}{M^2_\perp} \right )\, {\cal C}[ w^3_{LT}\, h_{1\sL}^{\perp\, q} \, h_{1\sT}^{\perp\,\bar q}]\,, \\
F_{TL}^{\,\cos (\phi_\sT - \phi_{S_1})} & = -  H \left ( z, \frac{M^2_\Q}{M^2_\perp} \right )\, {\cal C}[ w^1_{TL}\,  g_{1\sT}^{q}\, g_{1\sL}^{\bar q} \,] \,, \\
F_{TL}^{\,\cos (\phi_\sT + \phi_{S_1} - 2 \phi_\perp)} & = \Delta H \left ( z, \frac{M^2_\Q}{M^2_\perp} \right )\, {\cal C}[ w^2_{TL}\, h_{1}^{q}\,  h_{1\sL}^{\perp\, \bar q} \,]\,, \\
F_{TL}^{\,\cos (3\phi_\sT -\phi_{S_1} - 2\phi_\perp)} & = \Delta H \left ( z, \frac{M^2_\Q}{M^2_\perp} \right )\, {\cal C}[ w^3_{TL}\, h_{1\sL}^{\perp\, q} \, h_{1\sT}^{\perp\,\bar q}]\,, \\
 F_{TT}^{\, \cos(\phi_{S_1} -\phi_{S_2})} & = -H \left ( z, \frac{M^2_\Q}{M^2_\perp} \right )\,\bigg \{ {\cal C}[ w^1_{TT}\, f_{1\sT}^{\perp\, q} \, f_{1\sT}^{\perp\,\bar q}] \,+\, {\cal C}[ w^1_{TT}\, g_{1\sT}^{q} \, g_{1\sT}^{\bar q}]\bigg \}\,,\\
  F_{TT}^{\, \cos(2\phi_\sT -\phi_{S_1} -\phi_{S_2})} & = H \left ( z, \frac{M^2_\Q}{M^2_\perp} \right )\,\bigg \{ {\cal C}[ w^2_{TT}\, f_{1\sT}^{\perp\, q} \, f_{1\sT}^{\perp\,\bar q}] \,-\, {\cal C}[ w^2_{TT}\, g_{1\sT}^{q} \, g_{1\sT}^{\bar q}]\bigg \}  \,,
  \\
  F_{TT}^{\,\cos(\phi_{S_1} + \phi_{S_2}-2\phi_\perp)}& = \Delta H \left ( z, \frac{M^2_\Q}{M^2_\perp} \right )\, {\cal C}[ h_{1}^{q} \, h_{1}^{\bar q}] \,,\\
   F_{TT}^{\, cos(2\phi_\sT + \phi_{S_1} - \phi_{S_2}-2\phi_\perp)} & = \Delta H \left ( z, \frac{M^2_\Q}{M^2_\perp} \right )\, {\cal C}[ w_{TT}^3\, h_{1}^{q} \, h_{1\sT}^{\perp\,\bar q}] \,,\\
    F_{TT}^{\, cos(2\phi_\sT - \phi_{S_1} + \phi_{S_2}-2\phi_\perp)} & = \Delta H \left ( z, \frac{M^2_\Q}{M^2_\perp} \right )\, {\cal C}[ w_{TT}^4\, h_{1\sT}^{\perp\, q} \, h_{1}^{\bar q}] \,,\\  
    F_{TT}^{\, \cos(4\phi_\sT -\phi_{S_1} - \phi_{S_2} -2\phi_\perp)} & = \Delta H \left ( z, \frac{M^2_\Q}{M^2_\perp} \right )\, {\cal C}[ w_{TT}^5\, h_{1\sT}^{\perp\, q} \, h_{1\sT}^{\perp\,\bar q}] \,.
\end{align}
If we introduce the unit vector $\hat{\bm h} \equiv \bm q_\sT/ \vert \bm q_\sT \vert$, the transverse weights can be written as
\begin{align}
w_{UU} & =w_{LU} =w_{UL} = w_{LL}=\frac{2 (\hat {\bm h} \cdot \bm k_{1\sT}) (\hat{\bm h} \cdot \bm k_{2\sT}) - \bm k_{1\sT} \cdot \bm k_{2\sT}}{M_{h_1} M_{h_2} } = \frac{\vert \bm k_{1\sT}\vert \, \vert \bm k_{2\sT}\vert }{M_{h_1} M_{h_2}}\, \cos(2 \phi_\sT - \phi_1- \phi_2)\,, \label{eq:w2hh} \\
w^1_{TU} & = w^2_{UT} = w^2_{LT} = w^1_{TL} =\frac{\hat {\bm h} \cdot \bm k_{1\sT}}{M_{h_1}} =
\frac{\vert \bm k_{1\sT}\vert }{M_{h_1}}\,\cos(\phi_\sT - \phi_1)\,,\\
w^2_{TU} & = w^1_{UT} = w^1_{LT}  = w^2_{TL} = \frac{\hat {\bm h} \cdot \bm k_{2\sT}}{M_{h_2}} = \frac{\bm \vert \bm k_{2\sT}\vert }{M_{h_2}}\, \cos(\phi_\sT - \phi_2)\,,\label{eq:w_TU-2} \\
w^3_{TU} &  = w^3_{TL} =\frac{2(\hat {\bm h }\cdot \bm k_{1\sT})[2(\hat {\bm h }\cdot \bm k_{1\sT})(\hat {\bm h }\cdot \bm k_{2\sT})- \bm k_{1\sT} \cdot \bm k_{2\sT}] - \bm k_{1\sT}^2 (\hat {\bm h}\cdot \bm k_{2\sT})}{2\, M_{h_1^2}\, M_{h_2}}= \frac{1}{2}\frac{\bm k_{1\sT}^2 \,\vert \bm k_{2\sT}\vert}{M_{h_1}^2 M_{h_2}}\, \cos(2\phi_1 + \phi_2 - 3 \phi_\sT)\,,\\
w^3_{UT} & =  w^3_{LT}= \frac{2(\hat {\bm h }\cdot \bm k_{2\sT})[2(\hat {\bm h }\cdot \bm k_{1\sT})(\hat {\bm h }\cdot \bm k_{2\sT})- \bm k_{1\sT} \cdot \bm k_{2\sT}] - \bm k_{2\sT}^2 (\hat {\bm h}\cdot \bm k_{1\sT})}{2\, M_{h_1}\, M_{h_2}^2}=\frac{1}{2}\, \frac{\vert \bm k_{1\sT}\vert\,  \bm k_{2\sT}^2}{M_{h_1} M_{h_2}^2}\, \cos(\phi_1 + 2 \phi_2 - 3 \phi_\sT)\,, \label{eq:w3hh}\\
w^1_{TT} & = \frac{1}{2}\, \frac{\bm k_{1\sT} \cdot \bm k_{2\sT}}{M_{h_1} \,M_{h_2}} = \frac{1}{2}\, \frac{\vert \bm k_{1\sT} \vert \vert \bm k_{2\sT}\vert}{M_{h_1}\, M_{h_2}}\, \cos(\phi_1-\phi_2) \,, \\
w^2_{TT} & = \frac{2(\hat {\bm h }\cdot \bm k_{1\sT})(\hat {\bm h }\cdot \bm k_{2\sT})- \bm k_{1\sT} \cdot \bm k_{2\sT}}{2\, M_{h_1}\, M_{h_2}} = \frac{1}{2}\, \frac{\vert \bm k_{1\sT}\vert \vert \bm k_{2\sT}\vert}{M_{h_1}\,M_{h_2}}\, \cos(2\phi_\sT - \phi_1-\phi_2)\,,\\
w^3_{TT} & =\frac{2\, (\hat{\bm h}\cdot \bm k_{2\sT})^2 - \bm k_{2\sT}^2 }{2 M^2_{h_2}} = \frac{1}{2}\, \frac{\bm k^2_{2\sT}}{M_{h_2}^2}\, \cos 2 (\phi_\sT -\phi_2)\,,\\
w^4_{TT} & =\frac{2\, (\hat{\bm h}\cdot \bm k_{1\sT})^2 - \bm k_{1\sT}^2 }{2 M^2_{h_1}} = \frac{1}{2}\, \frac{\bm k^2_{1\sT}}{M_{h_1}^2}\, \cos 2 (\phi_\sT -\phi_1)\,,\\
w^5_{TT} & =  \frac{1}{4M_{h_1}^2M_{h_2}^2}\, \left \{ 2 \left[ 2 \, (\hat {\bm h} \cdot \bm k_{1\sT})\, (\hat {\bm h} \cdot \bm k_{2T}) - (\bm k_{1\sT} \cdot \bm k_{2\sT}) \right]^2 - \bm k_{1\sT}^2  \bm k_{2\sT}^2 \right \}  = \frac{1}{4}\, \frac{\bm k^2_{1\sT}\bm k^2_{2\sT}}{M_{h_1}^2M_{h_2}^2}\, \cos 2 (2 \phi_\sT - \phi_1 - \phi_2)\,.
\label{eq:w5_TT}
\end{align}

It is possible to single out the different angular modulations in Eq.~\eqref{eq:dsigma} by defining the azimuthal moments
\begin{align}
\langle {W(\phi_{S_1}, \phi_{S_2},\phi_\sT,\phi_\perp)}\rangle  & \equiv 2\,\frac {\int \d\phi_{S_1} \,\d\phi_{S_2} \,\d \phi_\sT \,\d\phi_\perp\, W(\phi_{S_1}, \phi_{S_2},\phi_\sT, \phi_\perp)\,\d\sigma(\phi_{S_1},\phi_{S_2},\,\phi_\sT,\,\phi_\perp)}{\int  \d\phi_{S_1} \,\d\phi_{S_2} \,\d \phi_\sT \,\d\phi_\perp\,\d\sigma( \phi_{S_1},\, \phi_{S_2} ,\,\phi_\sT,\,\phi_\perp)} \,.
\label{eq:mom-1}
\end{align}
When only one of the hadrons is transversely polarized, {\it e.g.} $S_{1 \sL} = S_{2\sL} = 0$, $\phi_{S_1}$ is integrated over and $\phi_{S_2}\equiv \phi_S$, one can also introduce the alternative observables:
\begin{align}
A_{UT}^{W(\phi_S,\phi_\sT, \phi_\perp)}  \equiv 
2\, \frac {\int \d \phi_\sT \,\d\phi_\perp\, W(\phi_S,\phi_\sT,\phi_\perp)\,\left [\d\sigma(\phi_S,\,\phi_\sT,\,\phi_\perp) - \d\sigma(\phi_S + \pi,\,\phi_\sT,\,\phi_\perp)\right ]}{\int \d \phi_\sT \,\d\phi_\perp\, \left [\d\sigma(\phi_S,\,\phi_\sT,\,\phi_\perp) + \d\sigma(\phi_S + \pi,\,\phi_\sT,\,\phi_\perp)\right ]} = \langle W(\phi_S,\phi_\sT, \phi_\perp) \rangle\,.
\label{eq:mom-2}
\end{align}

It is important to stress that the azimuthal modulations in Eq.~\eqref{eq:dsigma} and the corresponding TMD convolutions in Eqs.~\eqref{eq:F_UU}-\eqref{eq:w5_TT} are \emph{the same} as the ones obtained in Ref.~\cite{Arnold:2008kf} for the leading order channel $q\,\bar q \to \gamma^*\to \ell^+\ell^-$ of the Drell-Yan process $p\,p\to \ell^+\ell^-X$. We also note that the results in Ref.~\cite{Arnold:2008kf} are expressed in terms of the polar and azimuthal angles ($\theta,\phi$) of one of the decaying leptons in the Collins-Soper frame, where the virtual photon is at rest, with the additional choice $\phi_\sT=0$. As can be seen from Eq.~(57) of Ref.~\cite{Arnold:2008kf}, considering only the leading-twist terms, the cross section for the Drell-Yan process can be expressed in terms of two independent functions of the polar angle $\theta$, $(1+\cos^2\theta)$ and $\sin^2\theta$, multiplying an overall hard factor equal to one at ${\cal O}(\alpha_s^0)$. Similarly, only the two independent hard functions $H$ and $\Delta H$, given in Eq.~\eqref{eq:hard-f}, enter in the cross section for $p\,p\to \Q \Q\,X$.  However, $H$ and $\Delta H$ present a more involved analytical expression, in analogy to other quarkonium-related processes~\cite{denDunnen:2014kjo,Lansberg:2017dzg}.    

To conclude this section, we point out that single spin asymmetries have been studied  at tree level and assuming TMD factorization in $p\, p\to \pi \,\pi\,X$~\cite{Bacchetta:2005rm} and $p\, p\to \text{jet}\,\text{jet}\,X$~\cite{Bomhof:2007su} as well. Although the resulting azimuthal modulations are similar to the ones we find for $p\, p \to \Q \, \Q\, X$, as shown in Ref.~\cite{Rogers:2010dm}, TMD factorization is expected to be broken for the processes in Refs.~\cite{Bacchetta:2005rm,Bomhof:2007su}, because of the combined effects ISIs and FSIs~\cite{Rogers:2010dm}. On the other hand, as already mentioned in the Introduction, for the reaction under study, if the final quarkonia are produced directly in a color singlet state, only ISIs between the active partons and the spectators can occur. Hence the color-flow structure is similar to the Drell-Yan process, for which TMD factorization has been established.

\section{\label{sec:pheno} Phenomenology}

In this section we describe our phenomenological analysis for $q\bar q$-induced double quarkonium production in hadronic collisions, where the quarkonium states considered are the $J/\psi$, $\psi(2S)$ and $\Upsilon$ mesons. We first compare our results with COMPASS data on the unpolarized cross section for di-$J/\psi$ production~\cite{COMPASS:2022djq}. We then provide predictions for the following azimuthal moments defined via Eqs.~\eqref{eq:mom-1}-\eqref{eq:mom-2}:
\begin{align}
\langle \cos2(\phi_\sT - \phi_\perp) \rangle & = \frac{F_{UU}^{\cos{2(\phi_\sT - \phi_\perp)}}}{F_{UU}} = \frac{\Delta H}{H}\,\frac{{\cal C}[w_{UU}\,h_1^{\perp\, q}\, h_1^{\perp\, \bar q}]  }{{\cal C}[f_1^q\, f_1^{\bar q}]}\,, \label{eq:cos2phi} \\
\langle \sin(\phi_\sT + \phi_S - 2\phi_\perp) \rangle & = A_{UT}^{\sin{(\phi_\sT + \phi_S - 2\phi_\perp)}}  =   \frac{F_{UT}^{\sin{(\phi_\sT + \phi_S - 2\phi_\perp)}}}{F_{UU}} 
= \frac{\Delta H}{H}\,\frac{{\cal C}[w_{UT}^{2}\,h_1^{\perp\, q}\, h_1^{\bar q}] }{{\cal C}[f_1^q\, f_1^{\bar q}]}
\,, \label{eq:sin2phi_h1Tp_h1}\\
\langle \sin(\phi_\sT - \phi_S) \rangle & = A_{UT}^{\sin{(\phi_\sT - \phi_S)}}  = \frac{F_{UT}^{\sin{(\phi_T - \phi_S)}}}{F_{UU}}  = \frac{{\cal C} [w^1_{UT}\,  f_1^q\, f_{1\sT}^{\perp\, \bar q}\,]}{{\cal C}[f_1^q\, f_1^{\bar q}]}\,, 
\label{eq:AUT_Sivers}
\end{align}
where the hard factors $H$ and $\Delta H$ are given in Eq.~\eqref{eq:hard-f}, while the explicit expressions of the weight functions $w_{UU}, w_{UT}^2$ and $w^1_{UT}$ can be found in Eqs.~\eqref{eq:w2hh}-\eqref{eq:w_TU-2}. The azimuthal moments in Eqs.~\eqref{eq:sin2phi_h1Tp_h1}-\eqref{eq:AUT_Sivers} can  be defined when one of the incoming hadrons is transversely polarized. As we assume that the polarized hadron is always $h_2$, it is understood that $\phi_S \equiv \phi_{S_2}$.  The kinematic regions investigated in our study are the ones relevant for COMPASS/AMBER~\cite{Adams:2018pwt}, as well as  the present and future fixed-target experiments at the LHC, namely SMOG~\cite{Thesis_SMOG, LHCb:2014vhh},  SMOG2~\cite{BoenteGarcia:2024kba} and the LHCspin project~\cite{Aidala:2019pit,LHCspin:2025lvj}.

Concerning the (anti)quark TMDs probed in the above asymmetries, we point out that several extractions of unpolarized and polarized proton and pion TMDs have been performed by many groups at various perturbative orders and with different logarithmic accuracy~\cite{Anselmino:2013lza,Boglione:2018dqd,Boglione:2024dal,Bacchetta:2017gcc,Bacchetta:2019sam,Cerutti:2022lmb,Bacchetta:2024qre,Bacchetta:2025ara,Scimemi:2019cmh,Vladimirov:2019bfa,Moos:2023yfa,Moos:2025sal}. In the present analysis, we employ several TMD parameterizations by the MAP Collaboration through {\tt TMDlib}~\cite{Hautmann:2014kza,Abdulov:2021ivr}. For the pion we adopt the MAP\-TMD\-Pion22 TMDs at next-to-leading log (NLL) accuracy~\cite{Cerutti:2022lmb,Rossi:private}, while for the proton TMDs we use the PV17~\cite{Bacchetta:2017gcc} and MAP22~\cite{Bacchetta:2022awv} extractions at NLL. In particular, the PV17 parametrization has been chosen in order to  give consistent predictions for the Sivers asymmetry in Eq.~\eqref{eq:AUT_Sivers}, as it was taken as the baseline for the unpolarized cross section in the extraction of the PV20 quark Sivers function~\cite{Bacchetta:2020gko}  we use here. The computations involving the less-known quark Boer-Mulders function are performed by saturating its  positivity bound:
\begin{equation}
    \lvert h_1^{\perp q} (x, \bm{k}_\sT)\rvert \leq \frac{M_h}{\vert \bm k_\sT\vert} f_1^q(x, \bm{k}_\sT)\,,
    \label{eq:BM-pos-bound}
\end{equation}
where $M_h$ is the hadron mass. For the transversity distribution we adopt the extraction performed in Ref.~\cite{Boglione:2024dal}. In this way we estimate the upper bounds of $\langle \cos2(\phi_\sT - \phi_\perp) \rangle$, whereas in our predictions for $\langle \sin(\phi_\sT + \phi_S - 2\phi_\perp) \rangle$ we maximize only $h_1^{\perp q}$.  This different treatment of $h_1^{\perp\,q}$ as compared to the other TMDs investigated here is also motivated by the fact that the sign of $h_1^{\perp q}$ for different quark flavors is not yet determined from phenomenological analyses~\cite{Zhang:2008nu,Lu:2009ip,Barone:2009hw,Barone:2010gk,Wang:2018naw}.

The uncertainties on our results are the ones on the parameterizations of the TMDs that we use. The central values are calculated by taking the mean of the TMD replicas, while the error bands correspond to the $1\sigma$ confidence region\footnote{In the case of asymmetric collisions, {\it e.g.}~$\pi\, p$ scattering, we first compute  all the combinations of replicas for the pion and proton TMDs, then we take the mean of these combinations and determine the corresponding 1$\sigma$ confidence region.}. Moreover, when dealing with asymmetric collisions ({\it e.g.}~$\pi\, p$ scattering), we calculate the value of $\alpha_s$ at the hard scale of the process, usually denoted by $Q$, here identified with the invariant mass of the produced quarkonium pair, that is $\alpha_s(\MQQ)$, through {\tt LHAPDF}~\cite{Buckley:2014ana}, by selecting the value corresponding to the specific set of collinear parton distributions, in terms of which the various proton TMD extractions are expressed.

Another important aspect of the present study is the choice of the parameters relative to the quarkonium states. The selected values of the quarkonium mass $M_{\cal Q}$ and the square of the radial wave function at the origin $\lvert R_\Q (0)\rvert^2$ are summarized in Table~\ref{tab:mass-R0-pars}. 
The values of $\lvert R_\Q (0)\rvert^2$ are based on a power-law type potential~\cite{Eichten:1995ch}. We stress that this is an intermediate choice, as it is well known that the value of $\lvert R_\Q(0)\rvert^2$ may vary much depending on the adopted $Q\overline  Q$ potential (see Tables I, II and III of Ref.~\cite{Eichten:1995ch}) and on the QCD perturbative order used to fix it through the measured quarkonium decay width (see Table A.2 of Ref.~\cite{ColpaniSerri:2021bla}). 
\begin{table}[tbp]
\normalsize
    \centering
    \begin{tabular}{c c c c c}
    \toprule
        Quarkonium & ~ &$M_\Q$  [GeV] & ~ & $\lvert R_\Q (0)\rvert^2$  [GeV$^3$]\\
         \midrule\\[-2mm]
         $J/\psi$    & ~ & 3.0 & ~ & 1.0\\[2mm]
%         ~ & ~ & ~ & ~ & ~ \\
         $\psi (2S)$ & ~ & 3.7 & ~ & 0.56\\[2mm]
 %        ~ & ~ & ~ & ~ & ~ \\
         $\Upsilon$  & ~ & 9.5 & ~ & 4.59\\[1mm]
         \bottomrule
    \end{tabular}
    \caption{List of mass and $\lvert R_\Q (0)\rvert^2$ values adopted for our phenomenological study. The values of $\lvert R_\Q (0)\rvert^2$ are taken from Ref.~\cite{Eichten:1995ch} for a power-law type potential.}
    \label{tab:mass-R0-pars}
\end{table}

It is also important to underline that in the present analysis we have taken into account only single parton scattering. At moderate energies, the double-parton scattering mechanism, in which the two final state quarkonia are produced from two independent partonic subprocesses, is expected to be strongly suppressed~\cite{Lansberg:2015lva}. Especially at COMPASS, it has been estimated to be smaller than the measured background and should be about 8\% of the single parton scattering  contribution~\cite{COMPASS:2022djq,Koshkarev:2019crs}. Further similar analyses at the fixed-target experiments at LHCb are anyway needed. 
Moreover, we do not include here any feed-down contribution from higher states. Nonetheless we notice, for instance, that the $\psi(2S)$ feed-down to di-$J/\psi$ production has been estimated to be about 50\% in Ref.~\cite{Lansberg:2019adr}. For this reason, in what follows, one should bear in mind that the di-$J/\psi$ yield could be enhanced by this amount, especially at energies higher than COMPASS, typical of the LHC. 

Finally, we reaffirm that, in order for TMD factorization to be applicable to the reactions under study, the two quarkonia must be produced directly as color-singlet states. For this reason, we have explicitly checked that, at LO in $\alpha_s$, the dominant background coming for the mixed CS-CO channel is suppressed w.r.t.\ the CS-CS one in $q \,\bar q \to {\cal Q}\,{\cal Q}\,X$ for the fixed-target kinematics we will consider in the following. This check has been done using {\tt HELAC-Onia}~\cite{Shao:2012iz,Shao:2015vga} and adopting different sets of CO LDMEs~\cite{Butenschoen:2011yh,Sharma:2012dy,Chao:2012iv}. The CS-CO contributions turn out to be \emph{at most} $\sim 10\%$ (in the double $J/\psi$ case) of the CS-CS ones, thus confirming the expected ${\cal O}(v^3)$ suppression predicted by NRQCD~\cite{He:2015qya}.

\subsection{\label{sec:COMPASS} Double $J/\psi$ production at COMPASS}

The COMPASS Collaboration has measured the unpolarized cross section for double $J/\psi$ production in $\pi^- p$ scattering using a negative pion beam with momentum $p_{\rm beam} = 190$ GeV off a NH$_3$ target, resulting in a c.m.~energy $\sqrt{s} \simeq 18.9$ GeV~\cite{COMPASS:2022djq}. Data are provided as a function of $M_{\psi\psi}^2$, $q_\sT$, and the two variables
\begin{equation}
x^{\psi\psi}_\myparallel = \frac{p_{\rm L}^{\psi\psi}}{p_{\rm beam}}\,,\qquad 
\lvert \Delta x^{\psi\psi}_\myparallel\rvert = \lvert x^{\psi_1}_\myparallel - x^{\psi_2}_\myparallel \rvert\,,
\label{eq:xparallel}
\end{equation}
where $x^{\psi_i}_\myparallel = p_{\rm L}^{\psi_i}/ p_{\rm beam}$, with $i=1,2$,  $p^{\psi_i}_{\rm L}$ being the component of the momentum of the single $J/\psi$ meson along the beam direction and $p^{\psi\psi}_{\rm L} = p^{\psi_1}_{\rm L} + p^{\psi_2}_{\rm L}$. The expression of the cross section differential in these variables, as well as more details on COMPASS kinematics, can be found in Appendix~\ref{sec:appendix-COMPASS-kinematics}.

\begin{figure}[b]
    \centering
    \includegraphics[height=6.5cm, keepaspectratio]{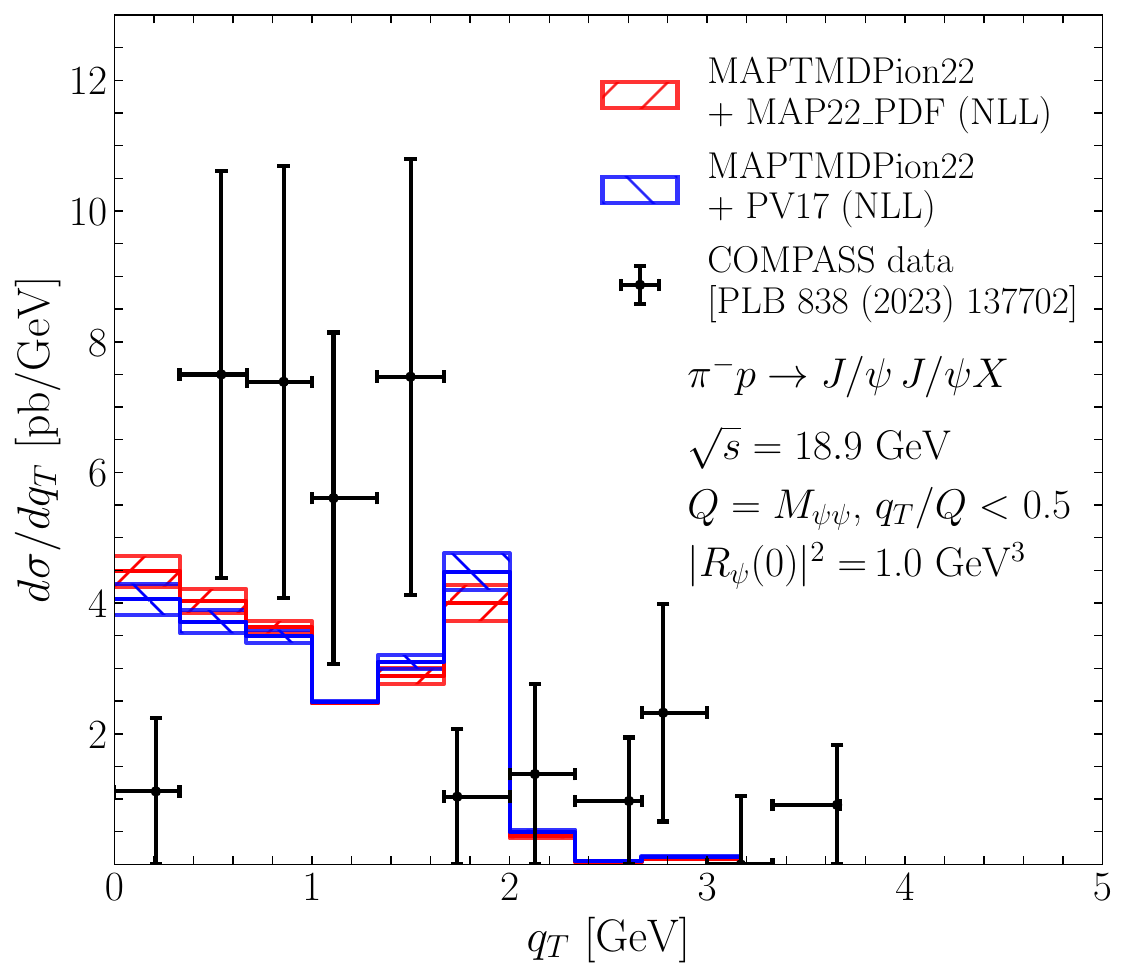} 
    \caption{Comparison with COMPASS data~\cite{COMPASS:2022djq} of the unpolarized cross section for di-$J/\psi$ production in $\pi^- p$ scattering, as a function of the transverse momentum of the $J/\psi$-pair, $q_\sT$, at $\sqrt{s} = 18.9$ GeV.}
    \label{fig:dsig-dqT-COMPASS}
\end{figure}

Our results for the unpolarized cross section for di-$J/\psi$ production are compared with the data~\cite{COMPASS:2022djq} in Fig.~\ref{fig:dsig-dqT-COMPASS}. We have estimated the gluon-induced channel directly, using the calculation in Ref.~\cite{Bor:2025ztq}, and found that it is negligible, of the order ${\cal O}(10^{-3})$ pb, within the kinematic region under investigation. Both predictions, based on the MAP22 and PV17 TMD parameterizations show a fairly good agreement with the data, considering that the COMPASS data sample is quite small: only about 25 di-$J/\psi$ events, see Table 2 of Ref.~\cite{COMPASS:2022djq}. While contributions from intrinsic charm and double-parton scattering are expected to be small at COMPASS~\cite{COMPASS:2022djq}, it remains to be seen whether a more precise determination of the radial wave function at the origin, the inclusion of next-to-leading order QCD corrections and feed-down effects can further improve our theoretical description. In our view, however, the present level of theoretical accuracy is sufficient for the purpose of estimating the order of magnitude of the unpolarized cross section and assess the experimental feasibility of the proposed measurements. Our main interest lies in the polarized TMDs and  the observables in Eqs.~\eqref{eq:cos2phi}-\eqref{eq:AUT_Sivers} for which} these uncertainties are supposed to cancel, at least partially, in the ratios.

\begin{figure}[t]
    \centering
\subfloat[\label{fig:cos2phi-sinphi-COMPASS}]{
    \includegraphics[height=6.5cm, keepaspectratio]{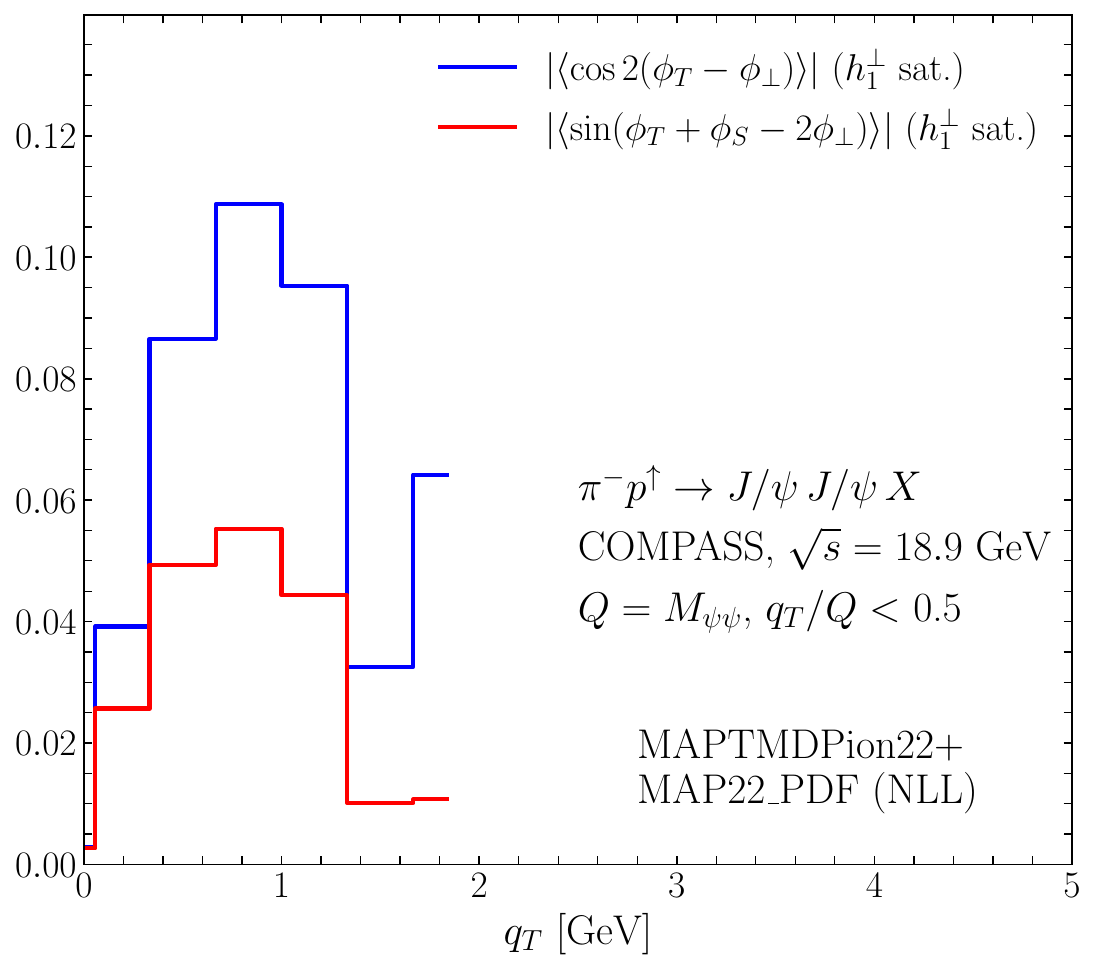} 
    }
\subfloat[\label{fig:AUT-Sivers-COMPASS}]{
\includegraphics[height=6.5cm, keepaspectratio]{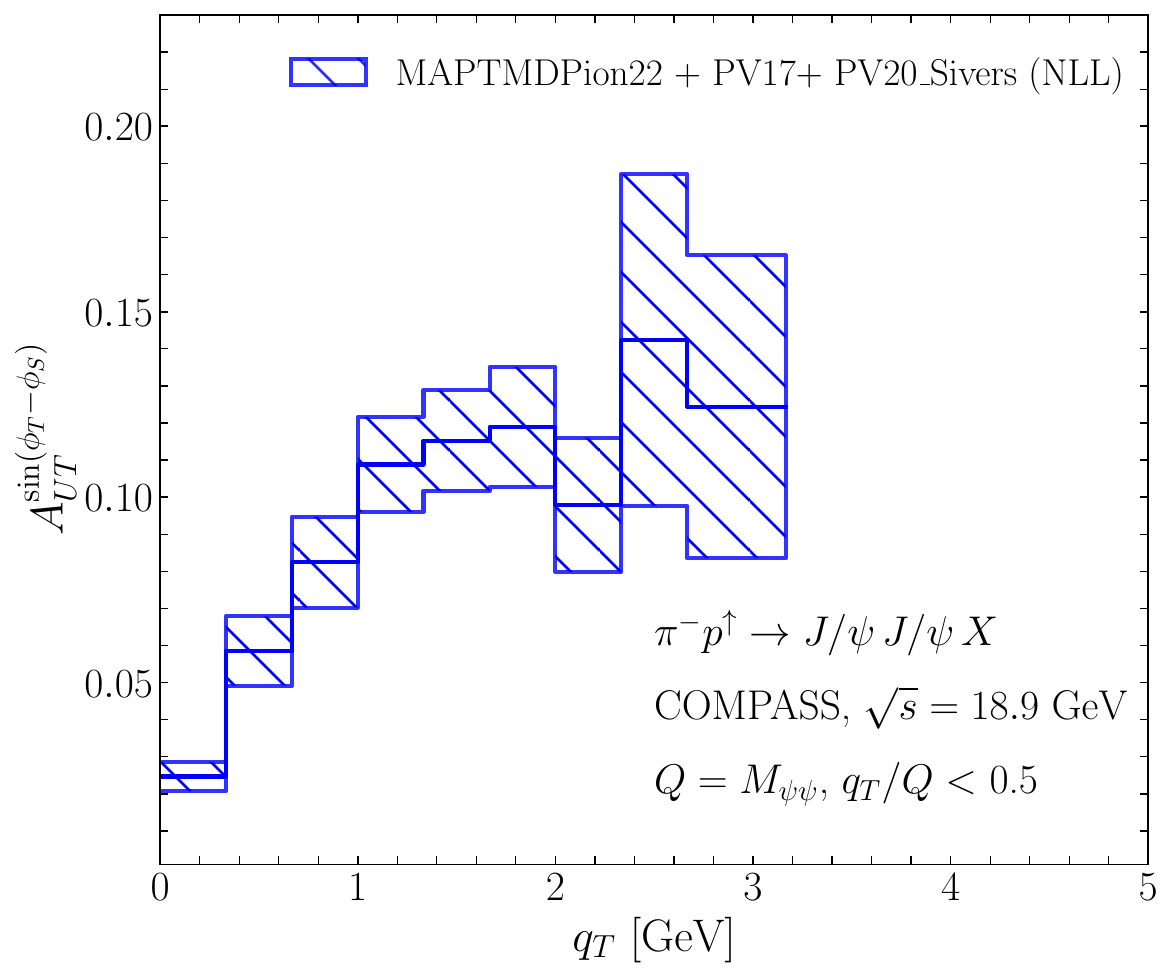}
    }
    \caption{Predictions for (a) $\lvert \langle \cos 2(\phi_\sT - \phi_\perp)\rangle \rvert$ and $\lvert \langle \sin(\phi_\sT + \phi_S - 2\phi_\perp)\rangle \rvert$, and (b) the Sivers asymmetry $\langle \sin(\phi_\sT - \phi_S )\rangle = A_{UT}^{\sin(\phi_\sT - \phi_S )}$ for di-$J/\psi$ production in $\pi^- p$ collisions at $\sqrt{s} = 18.9$ GeV at COMPASS. Computations are performed saturating the bound for $h_1^{\perp q}$ (see Eq.~\eqref{eq:BM-pos-bound}) and adopting the same $q_T$-binning as for the unpolarized cross section~\cite{COMPASS:2022djq}.}
    \label{fig:BM-h1Tph1-AUT-Siv-COMPASS}
\end{figure}

In Fig.~\ref{fig:BM-h1Tph1-AUT-Siv-COMPASS} we present our predictions, using the same kinematics and $q_\sT$-binning of the unpolarized cross section, for the $\langle \cos2(\phi_\sT - \phi_\perp) \rangle$, $\langle \sin(\phi_\sT + \phi_S - 2 \phi_\perp)\rangle$  (Fig.~\ref{fig:cos2phi-sinphi-COMPASS}) and the Sivers asymmetry $\langle \sin(\phi_\sT - \phi_S) \rangle$ (Fig.~\ref{fig:AUT-Sivers-COMPASS}). A maximum modulation of about 5-10\% and 2-5\% is estimated for $\lvert \langle \cos2(\phi_\sT - \phi_\perp) \rangle \rvert$ and $\lvert \langle \sin(\phi_\sT + \phi_S - 2 \phi_\perp) \rangle \rvert$ respectively. We note that the former is of the same order of magnitude as calculated for the Drell-Yan process~\cite{Wang:2018naw}. 
A sizable Sivers asymmetry (up to 10-15\%) is predicted in the same kinematic region. Notice that, as for the Drell-Yan process, we have assumed that $f_{1\sT}^{\perp \,q}$ has opposite sign w.r.t.~the Sivers function extracted from SIDIS. The sign of the asymmetry in Fig.~\ref{fig:AUT-Sivers-COMPASS} can be traced back to the dominant channel in the reaction ($\bar u u$), for which the valence region is picked up both in the pion and in the transversely polarized proton, with $f_{1\sT}^{\perp u}$ being positive in this process.

The data by COMPASS are very important but, as already mentioned above, limited in statistics. Repeating such a measurement at AMBER~\cite{Adams:2018pwt} would certainly help in enhancing the statistical precision of the current data.

\subsection{\label{sec:SMOG-LHCspin} Double quarkonium production at SMOG2 and LHCspin}

We present here our predictions for the unpolarized cross section, the $\lvert \langle \cos2(\phi_\sT - \phi_\perp) \rangle \rvert $, $\lvert \langle \sin(\phi_\sT + \phi_S - 2 \phi_\perp) \rangle \rvert $ and Sivers azimuthal asymmetries for the fixed-target programs at the LHC. Following what has been done at LHCb in the collider mode~\cite{LHCb:2016wuo, LHCb:2023ybt}, we show our results as a function of $q_\sT$ in different ranges of $\MQQ$, 
$\YQQ$ and $z$. In particular, we take $q_\sT / \MQQ < 0.5$  and $\YQQ \in [-1.5\!:\!0]$. 
The other kinematic cuts that have been imposed are summarized in Table~\ref{tab:kin-cuts-LHCb}. Using Eqs.~\eqref{eq:z1-z2}-\eqref{eq:Mpsipsi-MT-z}, it can be shown that the Jacobian for the change of variables $(z, \YQQ, \MQQ^2)  \mapsto (y_1, y_2, \bm K^2_\perp)$ is equal to one:
\begin{equation}
    \left\lvert \frac{\partial(z, \YQQ, \MQQ^2)}{\partial (y_1, y_2, \bm{K}^2_\perp)}\right\rvert = 1\,,
    \label{eq:LHCb-Jacobian}
\end{equation}
hence we have
\begin{equation}
    \frac{\d\sigma}{\d z\,\d\YQQ\, \d\MQQ^2 \,\d \phi_\perp \d^2\bm{q}_\sT} = \frac{\d\sigma}{\d y_1\, \d y_2\,  \d^2 \bm{K}_\perp\, \d^2 \bm{q}_\sT}\,,
\end{equation}
with the cross section on the right hand side given in Eqs.~\eqref{eq;dsigma-def}-\eqref{eq:dsigma}. An explicit derivation of Eq.~\eqref{eq:LHCb-Jacobian} is presented in Appendix~\ref{sec:appendix-LHCb-kinematics}. 
\begin{table}[b]
\normalsize
    \centering
    \begin{tabular}{c c c c c}
    \toprule
        Quarkonium & ~ & $M_{\Q\Q}$  [GeV] & ~& $z$ \\
         \midrule\\[-2mm]
         $J/\psi$  & ~ & $[7\!:\!10]$ & ~ & $[0.3\!:\!0.7]$\\[2mm]
%         ~ & ~ & ~ & ~ & ~ \\
         $\psi (2S)$ & ~ & $[9\!:\!12]$ & ~ & $[0.3\!:\!0.7]$\\[2mm]
 %        ~ & ~ & ~ & ~ & ~ \\
         $\Upsilon$  & ~ & $[20\!:\!30]$ & ~ & $[0.4\!:\!0.6]$\\[1mm]
         \bottomrule
    \end{tabular}
    \caption{Summary of the kinematic cuts used for the calculation of the unpolarized cross section at SMOG and LHCspin.}
    \label{tab:kin-cuts-LHCb}
\end{table}

%%%%%%%%%%%%%%%%%%%%%%%%%%%%%%%%%%%%%%%%%%%%%%%%%%%%%%%%%%%%%%%%%%%%%%%%%%%%%%%%%
\begin{figure}[t]
    \centering
\subfloat[\label{fig:dsig-dqT-psipsi}]{
    \hspace*{-2mm}\includegraphics[height=5.3cm, keepaspectratio]{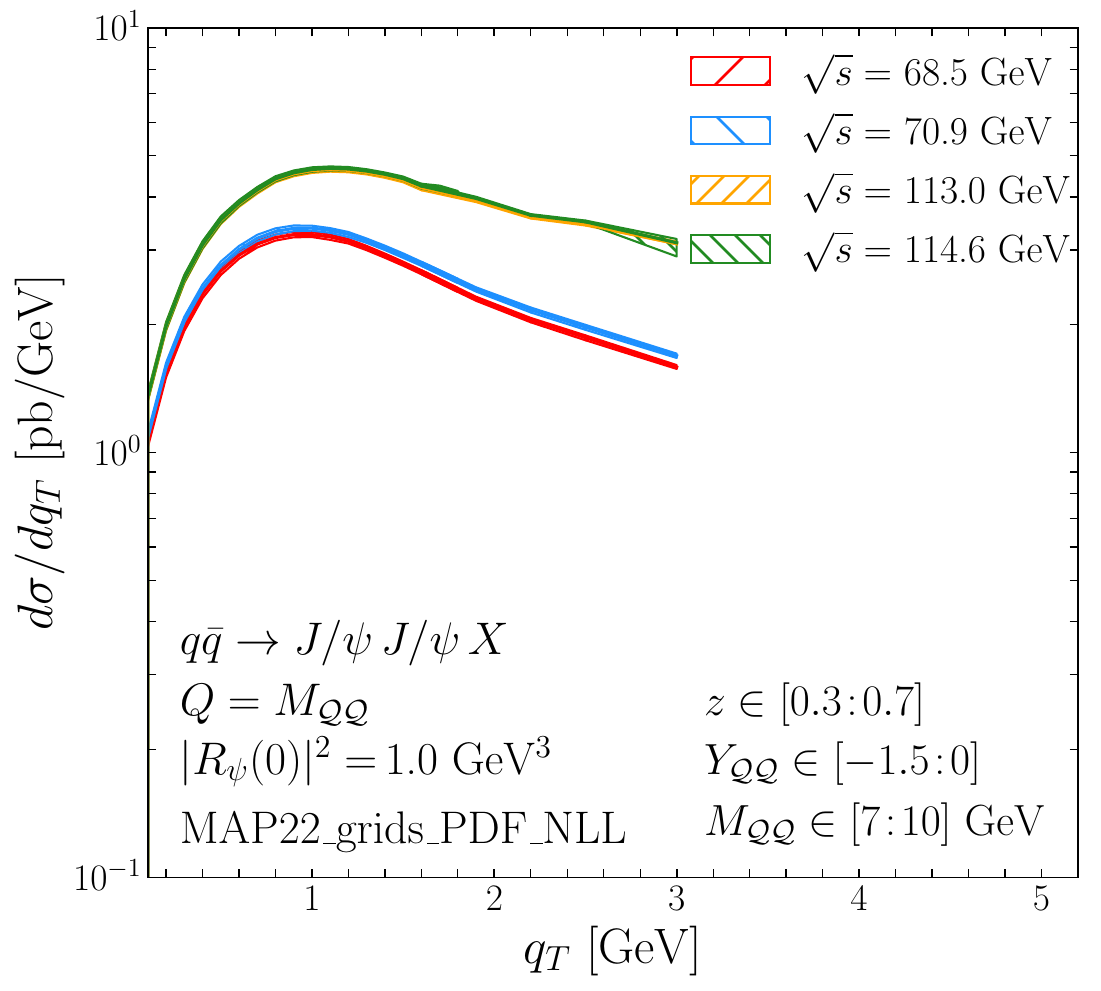}\hspace*{-2mm}
    }
\subfloat[\label{fig:dsig-dqT-psi2Spsi2S}]{    
    \includegraphics[height=5.3cm, keepaspectratio]{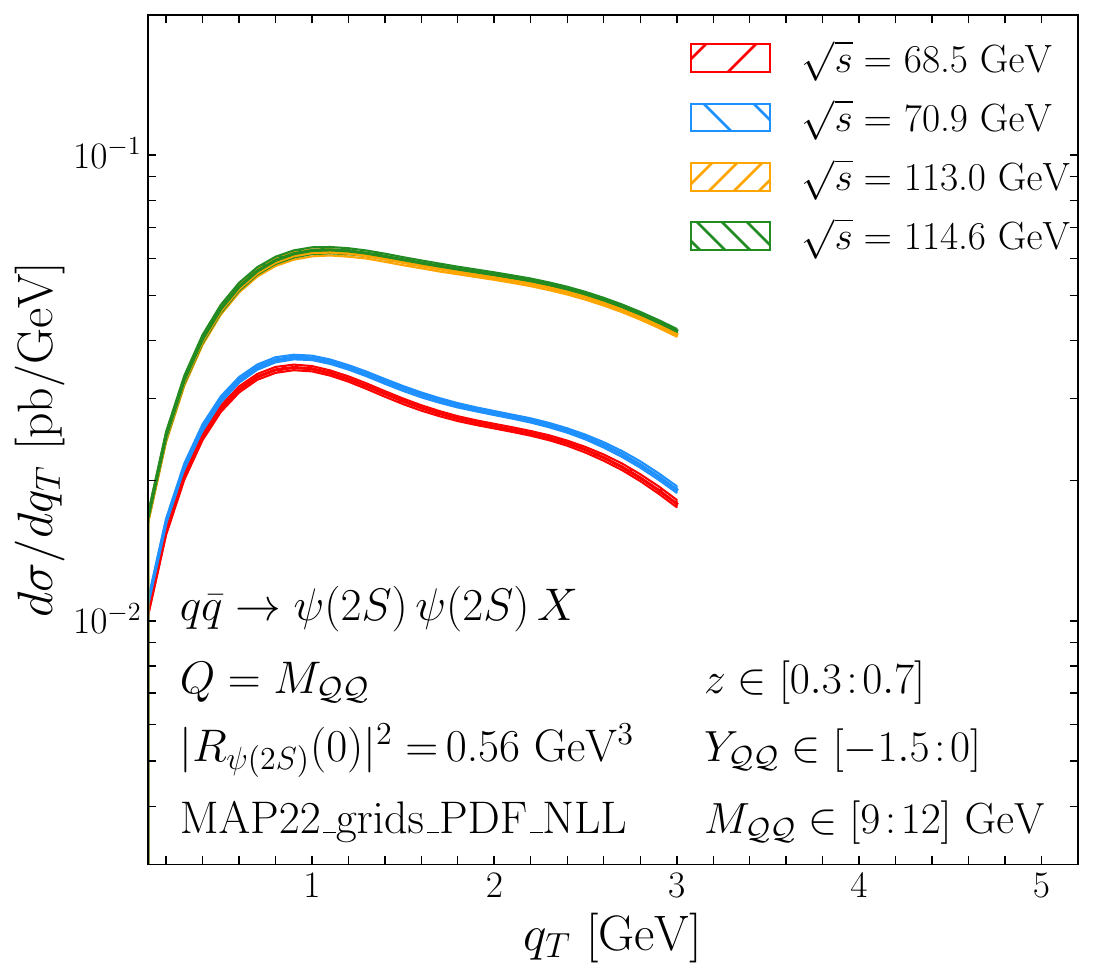}\hspace*{-2mm}
    }
\subfloat[\label{fig:dsig-dqT-Upsilon-Upsilon}]{
    \includegraphics[height=5.3cm, keepaspectratio]{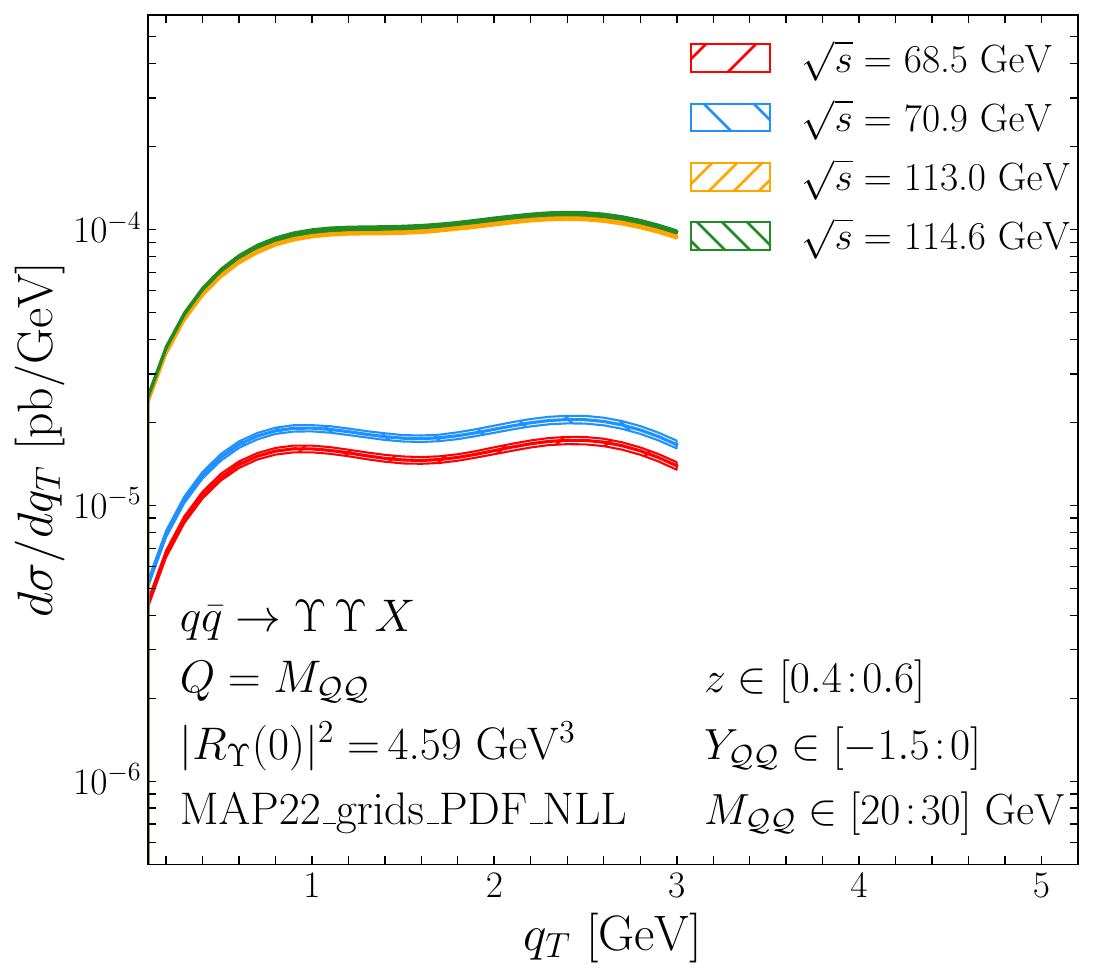}
    }
    \caption{Predictions for the quark induced unpolarized cross-section for (a) di-$J/\psi$, (b) di-$\psi(2S)$ and (c) di-$\Upsilon$ production at fixed-target experiments at LHCb for different values of $\sqrt{s}$ and in different $\YQQ$, $z$ and $\MQQ$ intervals. The MAP22 TMDs~\cite{Bacchetta:2022awv} have been employed in the computation.}
    \label{fig:dsig-dqT-LHCb-integr-pred}
\end{figure}

In Fig.~\ref{fig:dsig-dqT-LHCb-integr-pred} we gather our estimates of the unpolarized cross section for di-$J/\psi$ (Fig.~\ref{fig:dsig-dqT-psipsi}), di-$\psi(2S)$ (Fig.~\ref{fig:dsig-dqT-psi2Spsi2S}) and di-$\Upsilon$ (Fig.~\ref{fig:dsig-dqT-Upsilon-Upsilon}) production. Although not included in our plots, we have calculated explicitly the gluon-gluon fusion channel for the processes under study in the above-mentioned kinematical regions, using again the results of Ref.~\cite{Bor:2025ztq}. In contrast to what has been found for the COMPASS experiment, the gluon-gluon contribution is no longer negligible. It turns out to be at most 30-40\% of the quark-antiquark channel at $\sqrt{s} \approx 70$ GeV. Furthermore, it becomes a factor of two larger than the $q\bar q$ contribution at $\sqrt{s} \approx 115$ GeV. Once the so-far poorly known unpolarized gluon TMD is determined with a better accuracy, further dedicated studies will need to be performed to confirm these estimates.

With a projected integrated luminosity ${\cal L} ~\sim \mathcal{O}(100)$ pb$^{-1}$ for $p$H$_2$ collisions during the LHC Run 3~\cite{Bursche:2018orf}, considering the branching ratio ${\rm Br}(J/\psi \to \mu^+ \mu^-) = 5.961 \%$~\cite{ParticleDataGroup:2024cfk} and assuming that, compared to Run 2, the average di-$J/\psi$ detection efficiency for LHCb after the LHC upgrade for Run 3 will be higher than the present value $\langle \epsilon \rangle \sim 0.09$\footnote{This value is estimated on the basis of the number of events collected in Ref.~\cite{LHCb:2023ybt} for di-$J/\psi$ production at LHCb in the collider mode.}, the di-$J/\psi$ cross section we estimate in Fig.~\ref{fig:dsig-dqT-psipsi} should be measurable. 
Although we do not have similar information on the detection efficiency for di-$\psi(2S)$ and di-$\Upsilon$ production, it is reasonable to expect a similar improvement for these quarkonium states as well. One can also notice that the $q_\sT$-ranges we show for these quarkonia, especially for di-$\Upsilon$, could be extended to larger $q_\sT$ values within the bound $q_\sT / \MQQ < 0.5$. We refrain from doing so, as in this kinematic region one would probe the large-$x$ behavior of the TMDs, which is poorly constrained by the current extractions. Future measurements of di-$\Upsilon$ production would therefore be a valuable tool to constrain the unpolarized TMDs at large $x$.

\begin{figure}[t]
    \centering
\subfloat[\label{fig:AUT-Sivers-psipsi}]{
    \hspace*{-2mm}\includegraphics[height = 5.1cm, keepaspectratio]{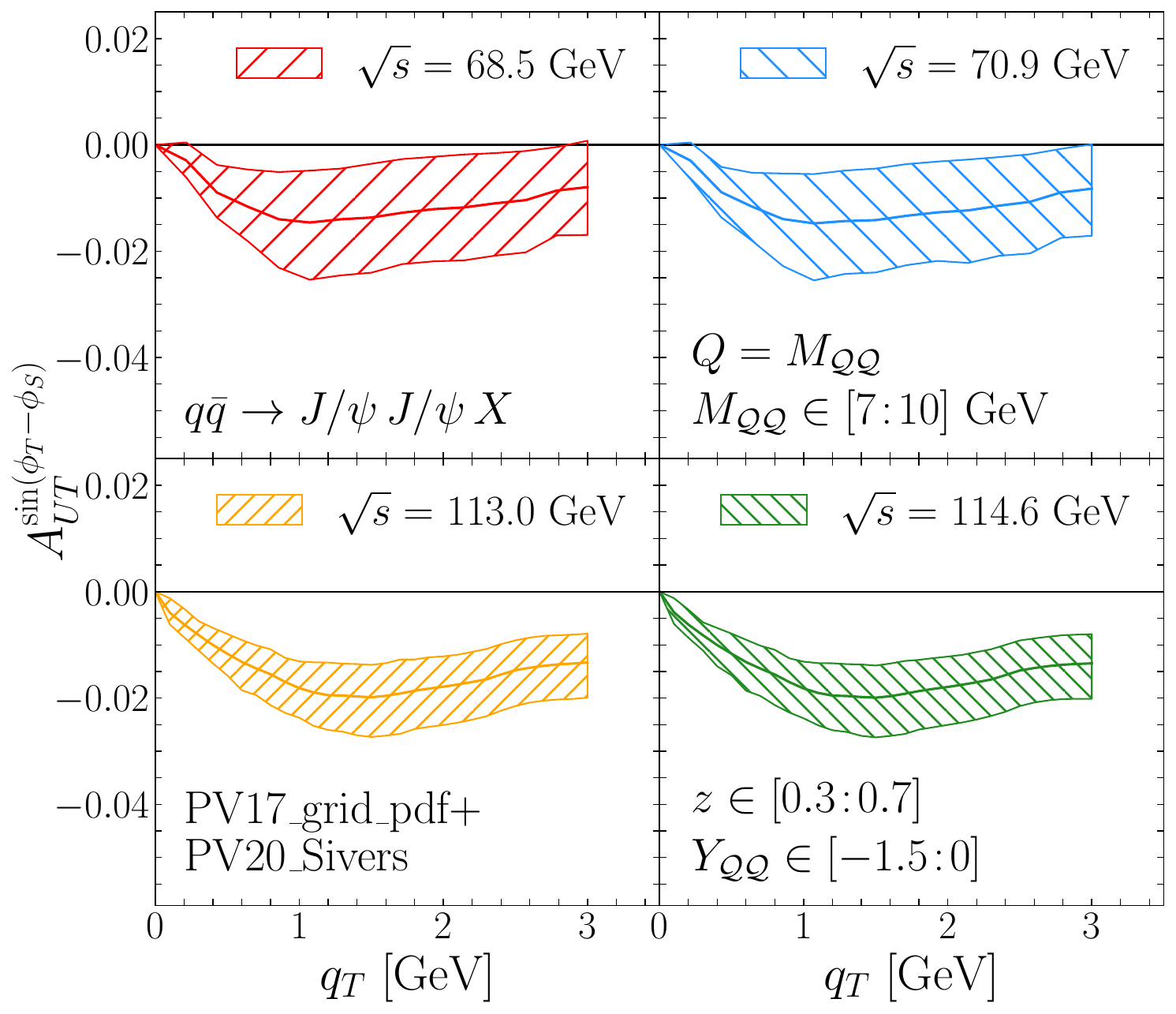}\hspace*{-2mm}
    }
\subfloat[\label{fig:AUT-Sivers-psi2Spsi2S}]{
    \includegraphics[height = 5.1cm, keepaspectratio]{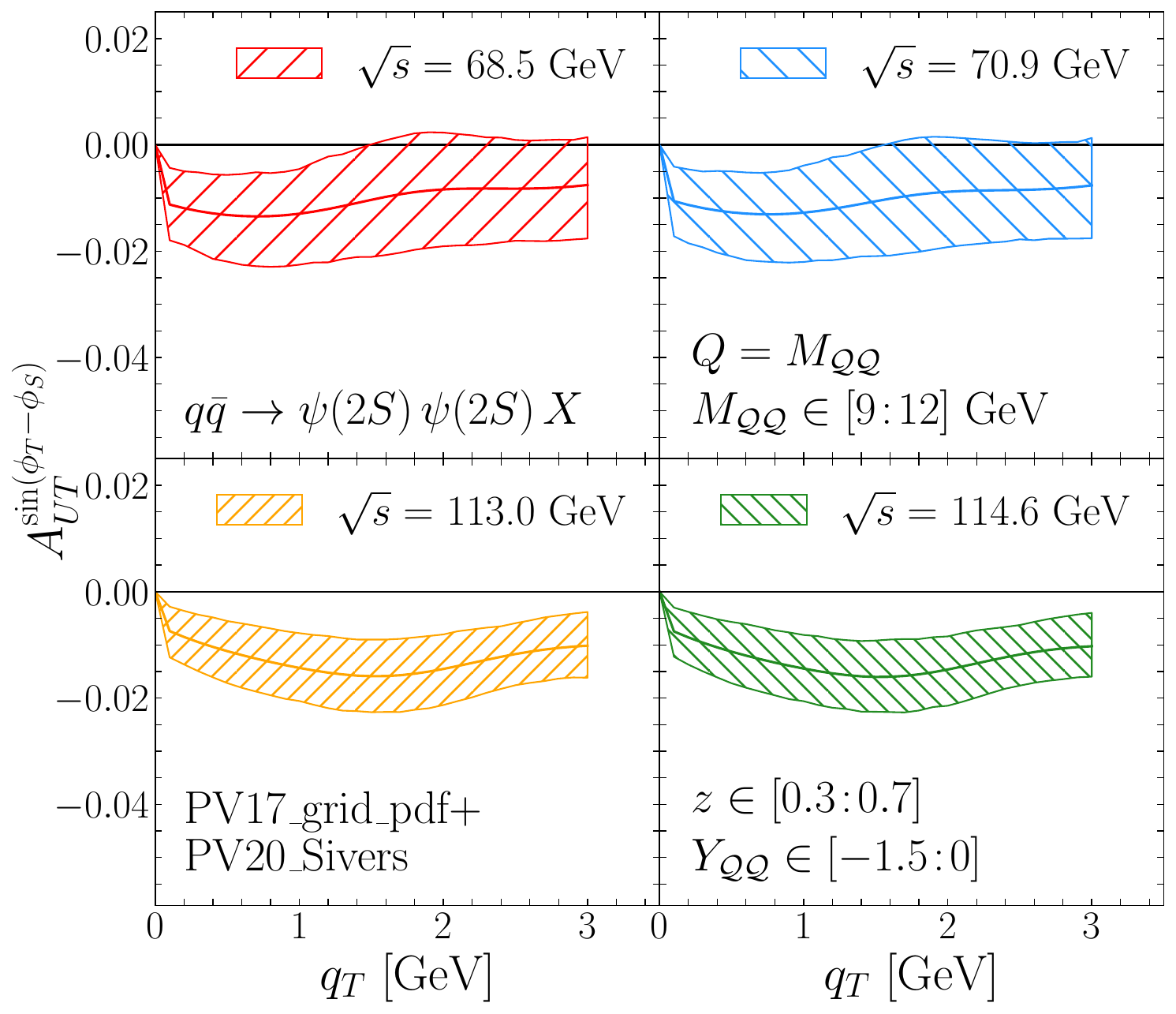}\hspace*{-2mm}
    }
\subfloat[\label{fig:AUT-Sivers-Upsilon-Upsilon}]{    
    \includegraphics[height = 5.1cm, keepaspectratio]{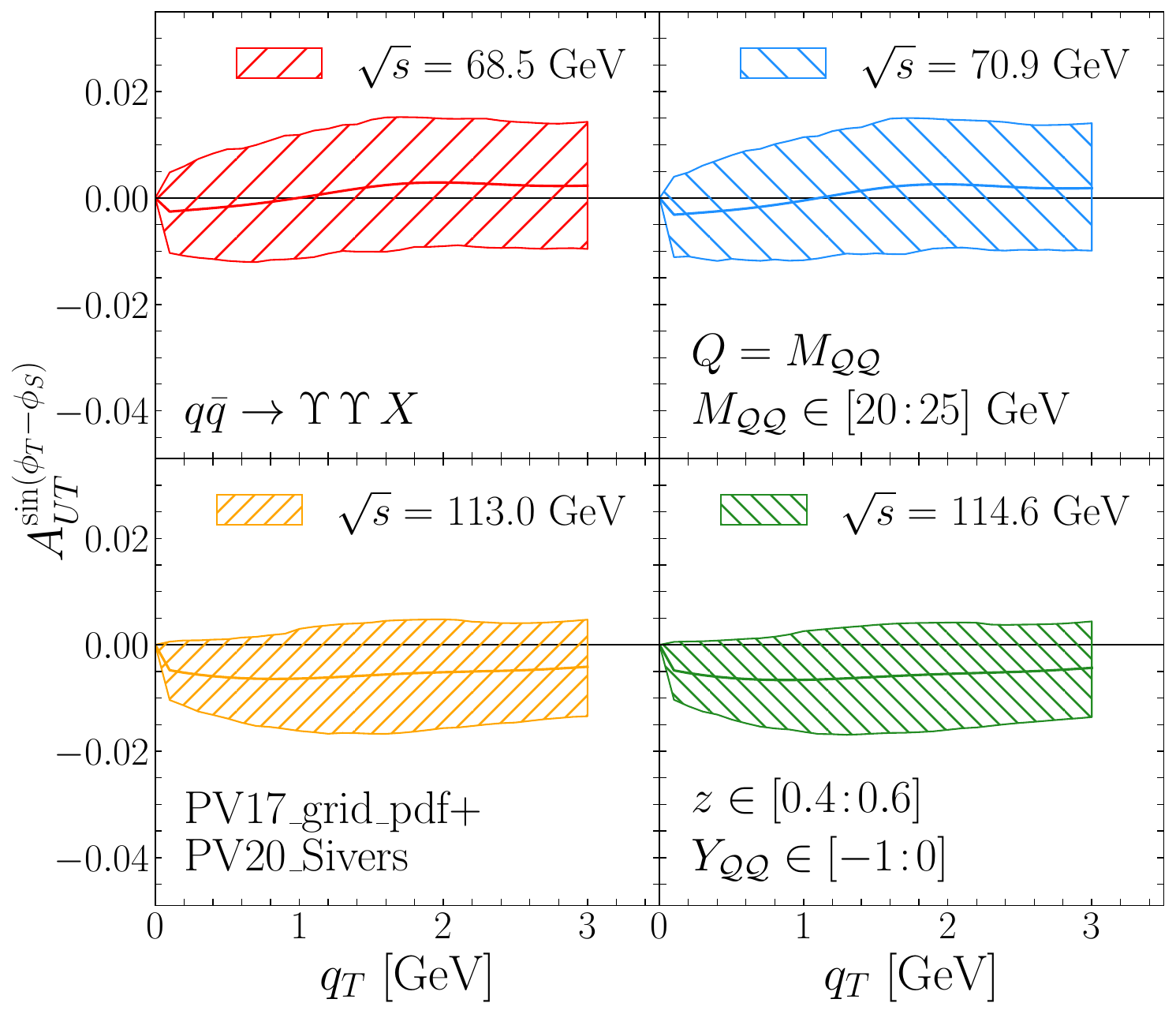}
    }
    \caption{Predictions for the quark induced Sivers asymmetry in double quarkonium production at fixed-target experiments at LHCb for different values of $\sqrt{s}$. Results are obtained  using the PV17~\cite{Bacchetta:2017gcc} and the PV20\_Sivers~\cite{Bacchetta:2020gko} parameterizations and assuming the sign change of the Sivers function with respect to SIDIS, as for the Drell–Yan processes.}
    \label{fig:AUT-Sivers-LHCb-pred}
\end{figure}

As polarized gas targets will be available at the planned LHCspin experiment, in Fig.~\ref{fig:AUT-Sivers-LHCb-pred} we present estimates for the Sivers asymmetry at different $\sqrt{s}$ values. Once again, we assume the sign change of the Sivers function with respect to the SIDIS process. A negative and nonzero (about 1-2\%) asymmetry is expected for di-$J/\psi$ (Fig.~\ref{fig:AUT-Sivers-psipsi}) and di-$\psi(2S)$ (Fig.~\ref{fig:AUT-Sivers-psi2Spsi2S}) production, while in the case of di-$\Upsilon$ (Fig.~\ref{fig:AUT-Sivers-Upsilon-Upsilon}) the asymmetry is compatible with zero, although with larger uncertainties. This can be traced back to the different kinematic range in $x_2$ explored for di-$\Upsilon$, as larger values of $x_2$ are probed where the magnitude of $f_{1\sT}^\perp$ is suppressed.  Moreover, at variance with $\pi^- p$ scattering at COMPASS, where only the $\bar u u$ contribution dominates, in $pp$ collisions other $q\bar q$ channels become relevant, {\it e.g.}~the ones involving the $d$-quark or the sea quark Sivers functions. As the latter are coupled with valence quarks from the unpolarized proton, and have opposite sign w.r.t.\ the one for the $u$-flavor, the asymmetry turns out to be negative. We emphasize that studies of the process dependence of TMDs, like the one proposed here, involving also a scrutiny of the Sivers sign change issue, are highly desirable and would certainly be possible at LHCspin. 

\begin{figure}[b]
    \centering
    \includegraphics[height = 7cm, keepaspectratio]{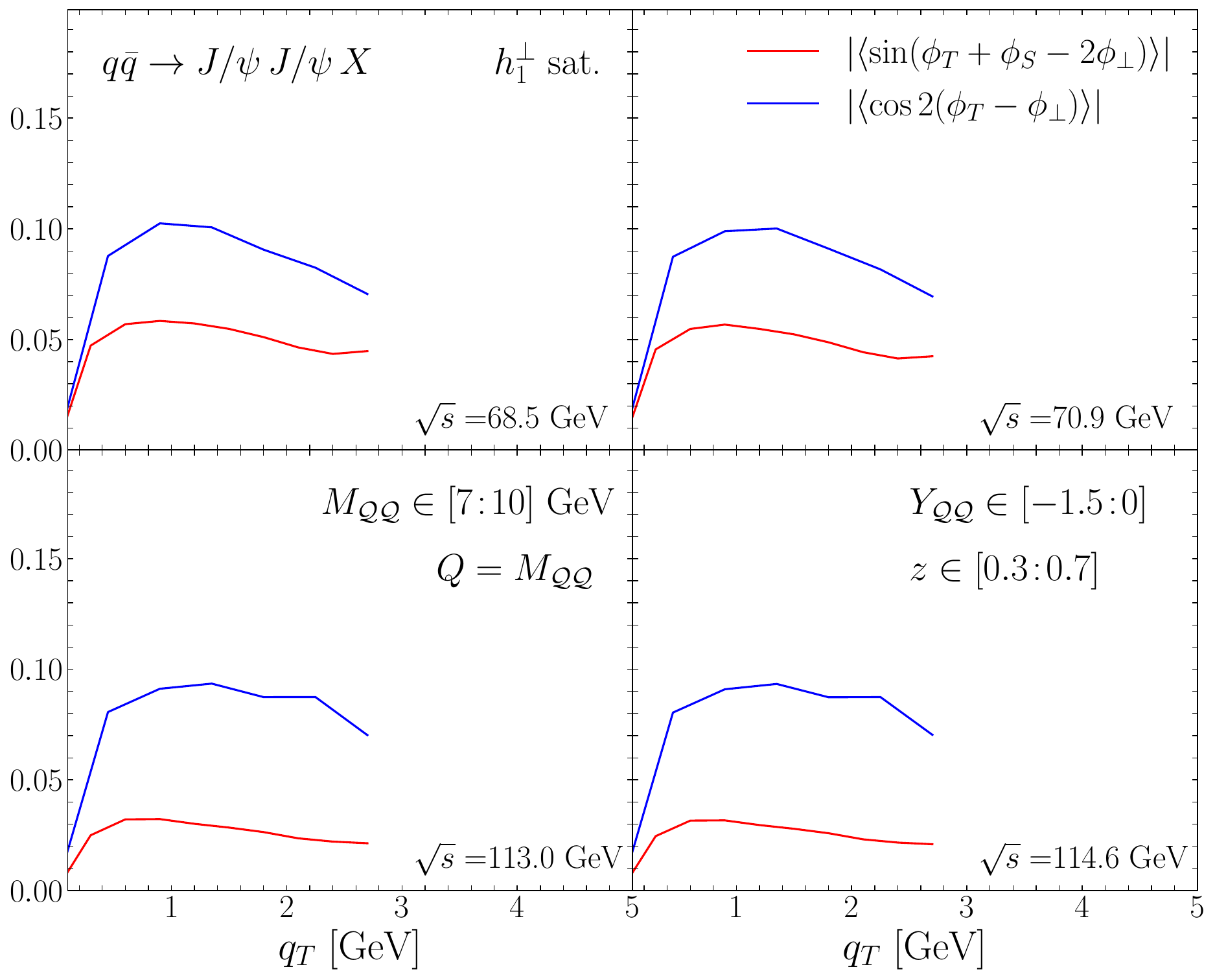}
    \caption{Predictions for the azimuthal asymmetries in Eqs.~\eqref{eq:cos2phi}-\eqref{eq:sin2phi_h1Tp_h1} for di-$J/\psi$ production at fixed-target experiments at LHCb for different values of $\sqrt{s}$. Results are obtained by saturating the positivity bound of the Boer-Mulders function $h_1^{\perp\,g}$ in Eq.~\eqref{eq:BM-pos-bound}, using the MAP22 unpolarized TMDs~\cite{Bacchetta:2022awv} and the transversity distribution $h_1^q$ extracted in Ref.~\cite{Boglione:2024dal}.}
    \label{fig:cos2phi-sin2phi-LHCb-pred}
\end{figure}

Finally, it is interesting to estimate the potential impact of transversely polarized quarks, {\it i.e.}~the asymmetries involving the Boer-Mulders function $h_1^{\perp q}$ and the transversity distribution $h_1^q$. In Fig.~\ref{fig:cos2phi-sin2phi-LHCb-pred} we present the predictions for the azimuthal asymmetries in Eq.~\eqref{eq:cos2phi}-\eqref{eq:sin2phi_h1Tp_h1} in di-$J/\psi$ production processes. As in the case of COMPASS (Fig.~\ref{fig:cos2phi-sinphi-COMPASS}), the maximized  $\lvert \langle \cos2(\phi_\sT - \phi_\perp) \rangle \rvert $ asymmetry is larger, ${\cal O} (10\%)$, than the $\lvert \langle \sin(\phi_\sT + \phi_S  - 2 \phi_\perp) \rangle \rvert$ ($\sim 5 \%$), which has to be expected since only $h_1^{\perp\, q}$, and not $h_1^q$, is maximized in our analysis. A slight increase (decrease) of the  $\lvert \langle \cos2(\phi_\sT - \phi_\perp) \rangle \rvert $ ($\lvert \langle \sin(\phi_\sT + \phi_S - 2\phi_\perp) \rangle \rvert $) asymmetry is observed at higher values of $\sqrt{s}$. Larger asymmetries, of the order of 10-20\% (not shown here) are expected at larger values of $\MQQ$ within the same $\YQQ$ and $z$ ranges, as well as for di-$\psi(2S)$ and di-$\Upsilon$ production. 

\section{\label{sec:conclusions} Summary and conclusions}

In this work, we have studied double quarkonium production in (un)polarized hadron-hadron collisions within the TMD factorization framework. Supported by NRQCD arguments, we have adopted the Color-Singlet Model to describe the quarkonium formation mechanism. We have derived the analytical expressions of the azimuthal modulations of the cross section generated in the quark-antiquark annihilation channel, showing that they can be expressed as convolutions of different leading-twist quark TMDs, in strong analogy with the well-known angular structure of the cross section of the Drell-Yan process.

Using very recent parameterizations of the unpolarized quark TMDs, we have found a fairly good agreement with COMPASS data on the unpolarized cross section for di-$J/\psi$ production in $\pi^- p$ scattering. For the same experiment, we have predicted sizeable azimuthal asymmetries, of the order of $5-10\%$, generated by the Boer-Mulders, transversity and Sivers quark TMDs. In particular, the large Sivers asymmetry ($10-15\%$) is dominated  by the $\bar u u$-channel, with the $\bar u$-quark being a valence quark for the $\pi^-$. Since the gluon contribution is strongly suppressed in di-$J/\psi$ production at COMPASS,  these measurements are ideal to probe the quark TMDs and test their universality properties, along with the already widely explored Drell-Yan and SIDIS processes. Future studies at AMBER will certainly help in improving the statistics collected so far at COMPASS.

Moving to the present and planned fixed-target experiments at LHCb, we have found that the $q\bar q$ annihilation channel in $p p \to J/\psi\, J/\psi\, X$ should remain dominant w.r.t.\ $gg$ fusion up to  $\sqrt{s} \approx 70$ GeV, whereas it should become subleading at larger c.m.~energies. Furthermore, a very small and negative Sivers asymmetry of the order of 1-2\% is estimated for di-$J/\psi$ and di-$\psi(2S)$ production because, in contrast to $\pi^-p$ collisions, in $pp$ collisions the $d-$ and sea-quark Sivers functions contribute as well, with opposite sign w.r.t.~the one for the $u$-flavor. A measurement of a larger Sivers asymmetry for double $J/\psi$ or $\psi(2S)$ production would therefore signal the presence of a nonzero Sivers function for gluons.
The Sivers asymmetry for di-$\Upsilon$ production turns out to be even smaller, but with larger uncertainties due to the large $x$-region probed in this case, where the TMDs are not well determined. Dedicated studies at LHCspin can thus help in constraining the quark Sivers function in a complementary kinematic region w.r.t.~the ones accessed in SIDIS, that is the process currently used to extract $f_{1\sT}^{\perp q}$. 
Similarly, by maximizing the Boer-Mulders function $h_1^{\perp\, q}$, we have predicted sizable $\lvert \langle \cos2(\phi_\sT - \phi_\perp) \rangle \rvert $ and $\lvert \langle \sin(\phi_\sT + \phi_S  - 2 \phi_\perp) \rangle \rvert$ asymmetries.  Measurements of these observables can shed light on the size and sign of the poorly known $h_1^{\perp\, q}$ distribution. Moreover, they could in principle allow to disentangle the quark and gluon contributions to the $\langle \cos2(\phi_\sT - \phi_\perp)\rangle$ modulation, corresponding to the TMD convolutions ${\cal C}[h_1^{\perp \,q}\,h_1^{\perp \,\bar q}]$ and ${\cal C}[f_1^{g}\,h_1^{\perp \,g}]$, respectively~\cite{Lansberg:2017dzg}.

Finally, we emphasize that our analysis of the $q\bar q$ annihilation channel in double quarkonium production in hadronic collisions complements ongoing theoretical and experimental studies, where only the $gg$ fusion channel has been taken into account. While the latter is certainly predominant at the LHC energies in the collider mode, we have shown that the quark contribution cannot be ignored in the fixed-target experiments at the LHC and becomes the leading one at the lower energies available at the COMPASS and AMBER experiments at CERN. We believe that a consistent description of the azimuthal asymmetries discussed in this paper, measured in double quarkonium production, SIDIS and the Drell-Yan processes at different energies, is necessary for a satisfactory knowledge of the proton TMDs, as well as their process dependence and QCD evolution properties. 

\section*{Acknowledgments}

This work is supported by the European Union “Next Generation EU” program through the Italian PRIN 2022 grant n.\ 20225ZHA7W. C.F. is further supported by the European Union’s Horizon Europe research and innovation programme under the Marie Skłodowska-Curie grant agreement n. 101150792 (STAT-TMDs). We are greatful to Dani\"el Boer, Alice Colpani Serri, Camilla De Angelis, Francesco Dettori, Federica Fabiano, Jean-Philippe Lansberg, Bakur Parsamyan, Luciano Libero Pappalardo and Lorenzo Rossi for useful discussions. We also thank Bakur Parsamyan for providing us with the COMPASS data, Filippo Delcarro and Lorenzo Rossi for providing us with the MAP22 pion and the PV20 Sivers TMD grids in the {\tt TMDlib} format. 

\appendix

\section{\label{sec:appendix-COMPASS-kinematics} COMPASS kinematics}

In this appendix we derive the general expression of the cross section for double $J/\psi$ production differential in the variables ${x^{\psi\psi}_\myparallel}$ and $\vert {\Delta x^{\psi\psi}_\myparallel}\vert$ adopted by the COMPASS Collaboration in Ref.~\cite{COMPASS:2022djq} and defined in Eq.~\eqref{eq:xparallel}.

To begin with, we notice that, as COMPASS is a fixed-target experiment, in the Sudakov decomposition of the initial pion and proton momenta given in Eq.~\eqref{eq:P1-P2}, we have $P_1^+ \approx \sqrt{2}\, p_\text{beam}$ and $P_2^- = M_p/\sqrt{2}$. Furthermore, in terms of the light-cone components of the outgoing quarkonia given in Eq.~\eqref{eq:K1-K2}, we can write
\begin{align}
x^{\psi\psi}_\myparallel = \frac{K_1^+ + K_2^+}{P_1^+} \left [1 - \frac{M_\perp^2}{2\, K_1^+K_2^+}  \right ]   \,, \qquad \Delta x^{\psi\psi}_\myparallel = \frac{K_1^+ - K_2^+}{P_1^+} \left [1 - \frac{M_\perp^2}{2\, K_1^+K_2^+}  \right ] \,.
\label{eq:x-Dx}
\end{align}
Similarly, the di-$J/\psi$ invariant mass is given by
\begin{align}
& M_{\psi\psi}^2  = (K_1 + K_2)^2 = K_1^2 + K_2^2 + 2K_1\cdot K_2 \approx 2M_{\cal Q}^2 + M_\perp^2 
\left ( \frac{K_1^+}{K_2^+} + \frac{K_2^+}{K_1^+} \right ) + 2 \bm K_\perp^2 
=   M_\perp^2 \, \frac{(K_1^+ + K_2^+)^2}{K_1^+ K_2^+}\,, 
\end{align}
where we have used the approximation $K_{1\perp} \approx - K_{2\perp} \approx K_\perp$ and $M_{1\perp} \approx M_{2\perp} \approx M_\perp$. Hence Eq.~\eqref{eq:x-Dx} becomes \begin{comment}
\begin{align}
 x^{\psi\psi}_\myparallel =  \frac{K_1^+ + K_2^+}{P_1^+}\, \left [1 - \frac{M^2_{\psi\psi}}{2(K_1^+ + K_2^+)^2} \right ]\, , \qquad  \Delta x^{\psi\psi}_\myparallel = \frac{K_1^+ - K_2^+}{P_1^+} \left [1 - \frac{M^2_{\psi\psi}}{2(K_1^+ + K_2^+)^2} \right ]\, ,
 \label{eq:x-Dx-2}
\end{align}
\end{comment}
\begin{align}
\label{eq:x-Dx-3}
  x^{\psi\psi}_\myparallel = \frac{{\cal K}_+}{P_1^+} \left ( 1 \, - \,  \frac{M^2_{\psi\psi}}{2\,{\cal K}_+^2} \right )\, , \qquad  \Delta x^{\psi\psi}_\myparallel = \frac{{\cal K}_-}{P_1^+} \left (1 - \frac{M^2_{\psi\psi}}{2\, {\cal K}_+^2} \right )\, ,  
\end{align}
with ${\cal K}_+ \equiv K_1^+ + K_2^+$ and ${\cal K}_- \equiv K_1^+ - K_2^+$
. The invariant phase space can then be written as
\begin{align}
\frac{\d^3 K_1}{2 E_1}\, \frac{\d^3 K_2}{2 E_2}  =  
\frac{\d K_1^+}{2 K_1^+}\,\frac{\d K_2^+}{2 K_2^+}\, \d^2{K_{1\sT}} \, \d^2{K_{2\sT}}
& = \pi \, \, \frac{\d K_1^+}{2 K_1^+}\,\frac{\d K_2^+}{2 K_2^+}\,\d  M^2_\perp \,  \d^2 q_\sT = \frac{\pi}{8 \, {\cal K}_+^2}\,\d {\cal K}_+ \, \d {\cal K}_- \, \d  M^2_{\psi\psi}\,  \d^2 q_\sT\, 
\label{eq:phsp}
\end{align}

In order to trade the variables $({\cal K}_+, {\cal K}_-)$  with $(x^{\psi\psi}_\myparallel, \Delta x^{\psi\psi}_\myparallel)$ we calculate the partial derivatives
\begin{align}
    \frac{\, \partial x^{\psi\psi}_\myparallel}{\partial \,{\cal K}_+} = \frac{1}{P_1^+} \left ( 1 \, + \,  \frac{M^2_{\psi\psi}}{2\,{\cal K}_+^2} \right )\, ,\quad     \frac{\, \partial \Delta x^{\psi\psi}_\myparallel}{\partial \,{\cal K}_+} = \frac{1}{P_1^+} \,  \frac{M^2_{\psi\psi}\,{\cal K}_- }{{\cal K}_+^3} \, ,\quad     \frac{\, \partial x^{\psi\psi}_\myparallel}{\partial \,{\cal K}_-} = 0\, ,\quad     \frac{\, \partial \Delta x^{\psi\psi}_\myparallel}{\partial \,{\cal K}_-} = \frac{1}{P_1^+} \, \left (1- \frac{M^2_{\psi\psi}}{2\, {\cal K}_+^2} \right ) \, ,
\end{align}
from which we obtain the Jacobian
\begin{align}
\left \vert \frac{\partial ({\cal K}_+, \, {\cal K}_-)}{\partial (x^{\psi\psi}_\myparallel,\, \Delta  x^{\psi\psi}_\myparallel) } \right \vert = \frac{4 \, (P_1^+)^2\, {\cal K}_+^4}{[4 \,{\cal K}_+^4 - M_{\psi\psi}^4]} = \frac{{\cal K}_+^2}{(x^{\psi\psi}_\myparallel) ^2\left [  1 +  \frac{M_{\psi\psi}^2}{x_\myparallel^{\psi\psi}  x_1 (P_1^+)^2}\right ]}\,.
\end{align}

Therefore the invariant phase space in Eq.~\eqref{eq:phsp} can be expressed as
\begin{align}
\frac{\d^3 K_1}{2 E_1}\, \frac{\d^3 K_2}{2 E_2}& = \frac{\pi}{8\,  (x^{\psi\psi}_\myparallel)^2 }\,\left [  1 +  \frac{M_{\psi\psi}^2}{2 \,x_\myparallel^{\psi\psi}  x_1 \, p_\text{beam}^2}\right ]^{-1} \d x_\myparallel^{\psi\psi} \, \d \Delta x_\myparallel^{\psi\psi} \, \d  M^2_{\psi\psi} \, \d^2 {\bm q}_\sT\,,
\label{eq:ps}
\end{align}

To express $x_{1,2}$ in terms of $x_\myparallel^{\psi\psi}$, one can start from Eq.~\eqref{eq:x-Dx-3} and write
\begin{align}
    \frac{{\cal K}_-}{P_1^+} - \frac{M_{\psi\psi}^2{\cal K}_-}{2 P_1^+ {\cal K}_+^2} - \Delta x_\myparallel^{\psi \psi} = 0 \,. 
\end{align}
Using $x_\myparallel^{\psi\psi} / \Delta x_\myparallel^{\psi\psi} = {\cal K}_+ / {\cal K}_-$, we get a second grade equation for ${\cal K}_+$:
\begin{align}
    {\cal K}_+^2 - x_\myparallel^{\psi\psi} P_1^+ {\cal K}_+ - \frac{M_{\psi\psi}^2}{2} = 0 \qquad \Rightarrow \qquad {\cal K}_+ = \frac{x_\myparallel^{\psi\psi} P_1^+}{2} \,\left(1 \pm \sqrt{1 + \frac{2M_{\psi\psi}^2}{(x_\myparallel^{\psi\psi} P_1^+)^2}}\right)\,.
\end{align}
The solution with the minus sign renders unphysical (negative) values for $x_1$, therefore the physical solution is the one with the plus sign. From Eq.~\eqref{eq:delta} we know that $x_1 = {\cal K}_+ / P_1^+$, and substituting the solution for ${\cal K}_+$ we get:
\begin{align}
x_1 = \frac{{\cal K}_+}{P_1^+} =\frac{x_\parallel^{\psi\psi}}{2}\left [1+ \left (1+ \frac{2 M_{\psi\psi}^2}{(x_\parallel^{\psi\psi}\,P_1^+)^2}\right )^{1/2}   \right] \,, \qquad x_2 =   \frac{M_{\psi\psi}^2}{2 P_2^-{\cal K}_+} = \frac{M_{\psi\psi}^2}{x_1\,s}\,.
\label{eq:x1-x2-COMPASS}
\end{align}
Numerically, the deviation of $x_1$ from $x_\myparallel^{\psi\psi}$ is of the order of per mille, hence the approximation $x_1 \approx x_\myparallel^{\psi\psi}$ would be accurate enough for the comparison with COMPASS data. Finally, one can write $z$ in terms of $x_\parallel, \Delta x_\parallel, M_{\psi\psi}$ as
\begin{align}
z = \frac{x_\parallel^{\psi\psi} - \Delta x_\parallel^{\psi\psi} }{2 x_1} \, \left [1- \frac{M^2_{\psi\psi}}{2 (x_1 P_1^+)^2}  \right ]\,.   
\end{align}
Using Eq.~\eqref{eq:delta}, re-expressing the invariant phase space in Eq.~\eqref{eq:PS} in terms of $x_\myparallel^{\psi \psi}$, $\vert \Delta x_\myparallel^{\psi \psi}\vert$ and $M_{\psi\psi}^2$, and after integrating over $x_1$ and $x_2$, the cross section in Eq.~\eqref{eq:PS} can be written in the final form
\begin{align}
\frac{\d\sigma}{ \d x_\myparallel^{\psi\psi} \, \d \lvert
\Delta x_\myparallel^{\psi\psi}\rvert
\, \d  M^2_{\psi\psi}\, \d^2 {\bm q}_\sT} & =   \frac{1}{s^2} \, \frac{1}{32 \pi}\, \frac{1}{(x_{\myparallel}^{\psi\psi})^2}\, \left[  1 +  \frac{M_{\psi\psi}^2}{2 \,x_\myparallel^{\psi\psi}  x_1 \, p_\text{beam}^2}\right ]^{-1}
\int \d^2 k_{1\sT}\, \d^2 k_{2 \sT} \, \delta^2(\bm k_{1\sT} + \bm k_{2\sT}-\bm q_\sT)\,\nonumber \\
& \qquad \times \sum_q\, \Phi^q(x_1, \bm k_{1\sT})  
\otimes \overline \Phi^q(x_2, \bm k_{2\sT})\,\otimes \vert {\cal M}_{q \bar q \to J/\psi J/\psi} \vert^2 \, + \{\Phi^q \leftrightarrow \overline{\Phi}^q \}\,,
\end{align}
with $x_1$ and $x_2$ given in Eq.~\eqref{eq:x1-x2-COMPASS}.

\section{\label{sec:appendix-LHCb-kinematics} LHCb kinematics}

We derive here explicitly the result in Eq.~\eqref{eq:LHCb-Jacobian}. We have to compute the Jacobian for the transformation $(y_1,\, y_2\, ,\bm{K}^2_\perp) \mapsto (z,\YQQ,\MQQ^2)$. It is easy to show that $\d^2 \bm{K}_\perp = \d\phi_\perp \d M_\perp^2$. From Eqs.~\eqref{eq:z1-z2}, \eqref{eq:z}, \eqref{eq:Ypsipsi} and \eqref{eq:Mpsipsi-MT-z}, we can write
\begin{align}
z = \frac{1}{1+e^{y_1-y_2}} \equiv \frac{1}{1+e^{\Delta y}}\,, \qquad 
\YQQ = \frac{y_1+y_2}{2} \,, \qquad
\MQQ^2 & =M_\perp^2\, \frac{(1+e^{\Delta y})^2}{e^{\Delta y}} \,,
\label{eq:lhcb-var}
\end{align}
where we have defined $\Delta y \equiv y_1 - y_2$. Hence, the Jacobian to be computed is
\begin{align}
 J=   \left\lvert \frac{\, \partial z \, \partial \YQQ \,\partial \MQQ^2}{\partial y_1\, \partial y_2 \,\partial M_\perp^2} \right\rvert \,= \,  \begin{vmatrix}
    - \frac{e^{\Delta y}}{(1 + e^{\Delta y})^2} & \frac{e^{\Delta y}}{(1 + e^{\Delta y})^2} & 0 \\
    \frac{1}{2} & \frac{1}{2} & 0 \\
    2 M_\perp^2\sinh{\Delta y} & -2 M_\perp^2\sinh{\Delta y} & \frac{(1+\e^{\Delta y})^2}{\e^{\Delta y}}
    \end{vmatrix}\,.
    \label{eq:LHCb-Jac}
\end{align}
Using the first of Eqs.~\eqref{eq:lhcb-var}, it can be shown that 
\begin{align}
\frac{e^{\Delta y}}{(1 + e^{\Delta y})^2} = z (1-z)
\end{align}
and substituting the expression above in Eq.~\eqref{eq:LHCb-Jac} we get $J = 1$.

\bibliographystyle{unsrtnat}
\bibliography{references}

\end{document}